# Findings of the 2[nd] Photonics and Electronics Technology for Extreme Scale Computing Workgroup (REPETE)

Design Challenges for Socket Level Photonic I/O

This document summarizes the current state of the art in electrical and optical interconnect technologies for HPC and contains guidance on the major challenges that are faced by HPC interconnect which are in need of attention.

# CONTENTS

# ACKNOWLEDGEMENTS

We wish to acknowledge the effort of the members of the 2nd PETE working group listed below and thank them for their hard work and insight that is summarized in this document.

| | |
|---|---|
| Keren Bergman | Mounir Meghelli |
| Joshua Fryman | Paul Petruzzi |
| Karen Grutter | Edward Preisler |
| Timothy Horton | Jonathan Proesel |
| Zhihong Huang | Thomas Salter |
| Ben Lee | Clint Schow |
| Tony Lentine | Marc Taubenblatt |
| Di Liang | Edward White |
| Michal Lipson | Ming Wu |

# EXECUTIVE SUMMARY

In mid 2011, the Department of Energy and the Department of Defense jointly sponsored a workgroup composed of subject matter experts spanning industry, academia, and government in high performance computing (HPC) system interconnect. Over a span of approximately six months, this workgroup developed a consensus on technology trends and key technical challenges related to HPC system interconnect where progress will have a major and lasting impact. The findings of this workgroup were released in a March 2012 workshop and report describing the findings to a wide audience of USG researchers, end users and program managers, system vendors, researchers from industry, academia and senior DoD, DoE, and USG leadership (1). The findings of this workgroup had a significant impact on USG strategic R&D priorities in HPC and photonic interconnect solutions and helped to direct USG investment and engagement with vendors of photonic integrated solutions over the last 7 years.

The initial PETE workgroup studied HPC interconnect and estimated progress until approximately the 2016-2017 time frame. To inform research activities in HPC interconnect of strategic importance to the USG beyond 2018, In January of 2018, the DoD sponsored the 2$^{nd}$ Photonics and Electronics Technology for Extreme-scale Computing (REPETE) workgroup. REPETE investigated new challenges in the area of HPC interconnect inspired by technology trends and challenges of vital importance to USG stakeholders.

The REPETE Working group investigated two focus areas of interest to the USG:

1. Socket Level Photonic I/O - Understand how performance limitations imposed on cross chip and off chip I/O (input/output) by copper interconnect impact technology roadmaps for socket I/O supporting memory, network, and coherent multiprocessor interconnect. Explore likely photonic solutions for overcoming the limitations imposed by copper interconnect, estimate impact on socket performance / capability / design, and document R&D challenges that will need to be addressed.

2. Cryogenic Photonic I/O - In the context of Superconducting SFQ-based (single flux quantum) computing, explore the limitations and technical requirements of moving data in and out of a cryogenic environment. Understand the cost and latency of moving data to/from memory that is either at 77k or room temperature resulting from a lack of dense cryogenic memory. Finally, investigate the interchip BW requirements stemming from circuit densities for SFQ-based logic that is very low compared to CMOS, and therefore may require multi-chip assemblies and novel packaging techniques.

In the model of the original PETE working group, the REPETE working group was a collaboration of device, component, and system level researchers in high performance computing (HPC) system interconnect and system architecture. The working group spanned industry, academia, and government, in research, development, product, and technology investment areas. To study each of these topics, the REPETE workgroup was subdivided into two teams, each independently investigating one of the two REPETE focus areas.

The workgroup team focusing on current and future design challenges for socket level photonic I/O began discussing technical challenges and current state of applying photonics to off-chip I/O in April of 2019 through biweekly meetings that concluded in early September of 2019. The focus of these meetings was to discuss what technology exists for moving the conversion of electrical signaling to photonic signaling from the node (as is currently done) down to within the socket for off-chip I/O. It is the belief of the workgroup members that direct conversion of I/O to photonics at the chip periphery, introduces opportunities for increased bandwidth to the chip socket, increased power efficiency over the traditional approach of converting to photonics at the node or rack periphery, and will be vital for meeting the exponential off-die bandwidth growth projected for 2025 and beyond.

The REPETE workgroup addressed a wide range of topics related to the challenges associated with increase in bandwidth for off chip and socket level photonic I/O. Areas discussed include:

- I/O Requirements and Trends at the Compute Socket for 2025 and Beyond
- Current and Near-Term Copper Solutions for Off-Chip, Socket-Level Interconnect
- Current Photonic Solutions for Off-Chip or Socket-Level Interconnect
- Light Generation for Off-Chip, Socket-Level I/O
- Fabrication and Packaging of Photonic Integrated Circuits

- All-Photonic Switching Technology

- Simulation of Photonic Interconnects

Similar to the original 2012 PETE report, this report focused its evaluation of photonic technologies using critical metrics including: Energy per bit, Bandwidth Density, Cost, Latency, and Reliability. For a subset of these metrics, the REPETE working group identified nominal target ranges that would enable chip/socket level photonic I/O to be used in lieu of electrical I/O by 2025 (Table 1).

**Table 1:** Nominal target ranges for key metrics to enable socket level photonic I/O.

| Metric | Target Range |
|---|---|
| Energy / bit | .1 – 1 pJ/bit |
| Bandwidth / socket | 100 Tbps – 1 Pbps / socket |
| Cost / bit | $.01 - $.10 / bit |
| Latency | 5 – 100 ns |

The power allocated to a die is quickly approaching a ceiling due to economic limits on cooling and power costs. Simultaneously, bandwidth demands are increasing exponentially. No mature technology is currently available to address these exponential issues. The main objectives of the REPETE workgroup were to determine what the main obstacles to meeting these needs are and what must be addressed with technologies in development to develop feasible solutions for high-performance computing. The consensus findings, key conclusions and recommendations are detailed below. It is the hope of the REPETE workgroup that these findings will help guide future strategic investments by the US government research community and its partners.

- **Existing off-chip I/O solutions do not meet the exponential off-die bandwidth growth projected for 2025 and beyond.** [1]

    Socket I/O demands are on an exponential trajectory but die size and allocated power have been on a linear trajectory. Off-die bandwidth already exceeds 8 Tbps, and the bandwidth demands are expected to continue to follow this exponential growth trend. Similar trends are visible in the LAN and device bandwidth requirements. With the exponential increase in socket bandwidth, we are quickly approaching the point, around 30 Tbps, when the power required to support the I/O requirements will be greater than the total power that is supplied to each socket (2).

- **Electrical advancement is not at a standstill, but we are at the long tail in improvements.** [1]

    Electrical interconnects are continuing to improve, but they are limited in distance and reach and we are approaching their Shannon capacity limits for common electrical I/O (CEI) standard long reach (LR, ~1 m) links. Efficiency and latency are also becoming issues; even for the current CEI standard at 56 Gbps, the efficiency is 2.5 pJ/bit for very short reach (VSR, 10 cm) and requires forward error correction (FEC). This efficiency continues to drop for longer reach and higher bandwidths. Proposed improvements to these interconnects are marginal; speed and efficiency will likely not increase much more.

- **For off-chip I/O, photonics will soon achieve energy efficiency comparable to (maybe even better than) copper.** [1]

    From a purely energy-efficient perspective, state-of-the-art integrated photonic links are approaching electrical link efficiency for VSR (~10 cm or internode) at around 2-10 pJ/bit. Should the goals of the DARPA Photonics in the Package for Extreme Scalability (PIPES) program be reached, this efficiency would be improved to sub-pJ/bit levels (3) – competitive with copper for off-chip I/O including VSR or longer distances. However, there are many other factors involved in the choice of electrical

---

[1] This is a new conclusion, not covered in PETE2012.

versus optical technologies, including cost, reliability, latency, packaging, thermal robustness, and architectural changes.

- **Cost is a major roadblock to the adoption of photonic interconnects. [1]**

    To be competitive with copper at even the common electrical I/O (CEI) standard for long reach (LR, ~1 m), the cost needs to be ~$0.10/Gbps, but commercially-available optical interconnect technologies are closer to $1/Gbps. Companies are targeting incrementally lower-cost solutions in the next few years, and ARPA-E's ENLITENED program is targeting $0.10/Gbps (4), but it should be noted that this program and commercial companies are primarily focused on datacenter applications (not HPC). Latency, in particular, is a performance metric that is not seen as important to datacenters, but is critical for HPC applications.

- **Standards and infrastructure for PIC fabrication, packaging, and assembly are immature, contributing to the high cost of photonic interconnect. [2]**

    As a whole, the number of photonic integrated circuit (PIC) foundry services is growing, but there are many aspects that need further development to achieve broad applicability, reliability, and low cost comparable to the CMOS ecosystem. Areas ripe for technological advancement include yield analysis, in-line process controls and process monitoring techniques, and standardization of fiber-to-chip coupling and interfacing with packaging. Unlike electronics, the device fabrication cost is closely tied to the packaging, in-line and end-of-line test, making assembly and packaging major challenges for photonics links (even at USR, VSR distances). Of particular note is the fact that current techniques for fiber-to-chip coupling are not optically efficient or scalable, especially when considering automated packaging/assembly and reliability. This is especially important as optical port count increases (as in the case for high port count transmitter, receiver, and all-photonic switch designs). In short, there does not yet exist an industry-standard suite of PIC packaging and testing tools; as a result, assembly/packaging is the major challenge for photonic links.

---

[1] PETE2012 also reflected on this topic.
[2] Packaging and assembly issues were also identified in PETE2012.

- **Non-coherent photonics will be required to meet energy efficient goals. [1]**

    Coherent photonic solutions currently require significant increases in total link efficiency to be viable for widespread datacenter deployment. Thus, they do not appear to be competitive with copper or direct-detection for short-range (<50m) ultra-efficient links. For links up to 2km, the larger link budgets and higher spectral efficiency for coherent links may be significant advantages, provided the link power efficiency can be reduced to levels comparable to direct detection.

- **Comb lasers are vital to achieving efficient, parallel, high-channel-count off-chip I/O. [2]**

    The power efficiency demands of off-chip I/O are pushing development in high channel count DWDM links operating at slower, more power efficient speeds (~10-25 Gbps).There is a diverse set of technologies available for light generation in optical interconnects for HPC systems, however, the trend towards DWDM architectures and advanced modulation formats (QAM) points to external comb lasers as a critical technology. While comb lasers can in principle produce 100's of λ's of laser energy over ~100 nm of spectrum (albeit with 10's of dB in power variation), the current state of the art for useful devices with output power balanced (within a few dB) across the spectrum is 30 - 60 λ's (often in the ~1300 nm window) at 20 – 50 GHz spacing producing <1 mW/λ, with 200 λ's at 35 GHz likely in the near future. Wall-plug efficiency is currently limited to ~10% (compared to ~40% for best-in-class high-power single-wavelength lasers) due to the saturable absorber used for comb generation, and the requirement for thermal stabilization for precise λ control. For more challenging system architectures requiring >3 mW/λ (example: higher order modulation formats), the state of the art is currently 24 λ's, with a reasonable target of 36 λ's at 50 – 100 GHz spacing.

---

[1] PETE2012 also reflected on this topic.

[2] This is a technical refinement of a conclusion from PETE2012.

- **Development of photonic simulation tools is vital given the diversity of photonic off-chip interconnect and switching solutions.** [1]

    Many promising new technologies are in development, but no <u>standard</u> solutions have been settled upon for advancing to higher bandwidth and better efficiency. Given the lack of standardization, photonic simulation tools such as PhoenixSim that plug into the CMOS design flow, will be vital to optimize interconnect architectures over the large design space resulting from the multitude of on-chip modulators, detectors, light sources, and multiplexing solutions.

- **Photonic circuit switches may offer a more efficient switching technology but are not a drop-in replacement for existing electrical switches.** [1]

    Much like electrical interconnect technology, traditional electrical switching will soon be hitting a thermal wall as the need for speed increases. Photonic switches may offer a more power-efficient switch option, but no photonic technology provides a likely roadmap to bit-level switching. Photonic switching speeds and the lack of memory state are only sufficient for circuit switching, and thus are not a drop-in replacement for existing electrical switches. Further driving the need for device to system level photonic simulation tools, additional HPC research will need to develop network topologies, routing, and controls for circuit-switched, stateless (no memory buffers in the switch) routing in HPC systems.

Disclaimer: The results presented at REPETE and reflected in this document do not necessarily represent the intentions or opinions of the organizations represented. The material presented is derived from the literature or represents the opinions of the members of the REPETE working group.

---

[1] This is a new conclusion, not covered in PETE2012.

# TRENDS IN I/O REQUIREMENTS FOR 2025 AND BEYOND

Based on materials and presentation provided by Josh Fryman (Intel)

Socket I/O demands are on an exponential trajectory, but key parameters of socket performance with current technology will eventually fail to meet these demands. In order to determine a solution to this challenge, we must first understand the driving forces behind socket I/O demands.

We look to past growth trends to predict where performance is going. At the die level, die size and allocated power have been on a linear trajectory (see Figure 1). The linear growth of die power is primarily limited from an economic perspective; it is possible for dies to handle higher thermal loads through various engineering solutions but paying for those solutions or moving to larger dies is too expensive.

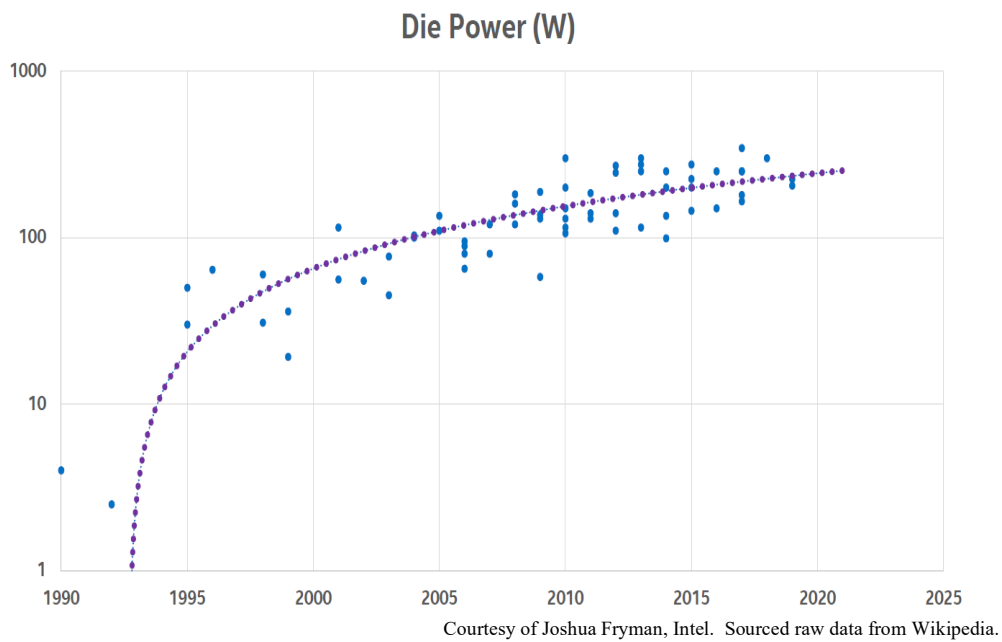

Courtesy of Joshua Fryman, Intel. Sourced raw data from Wikipedia.

**Figure 1**: Die power growth history. Purple dotted line is a linear fit.

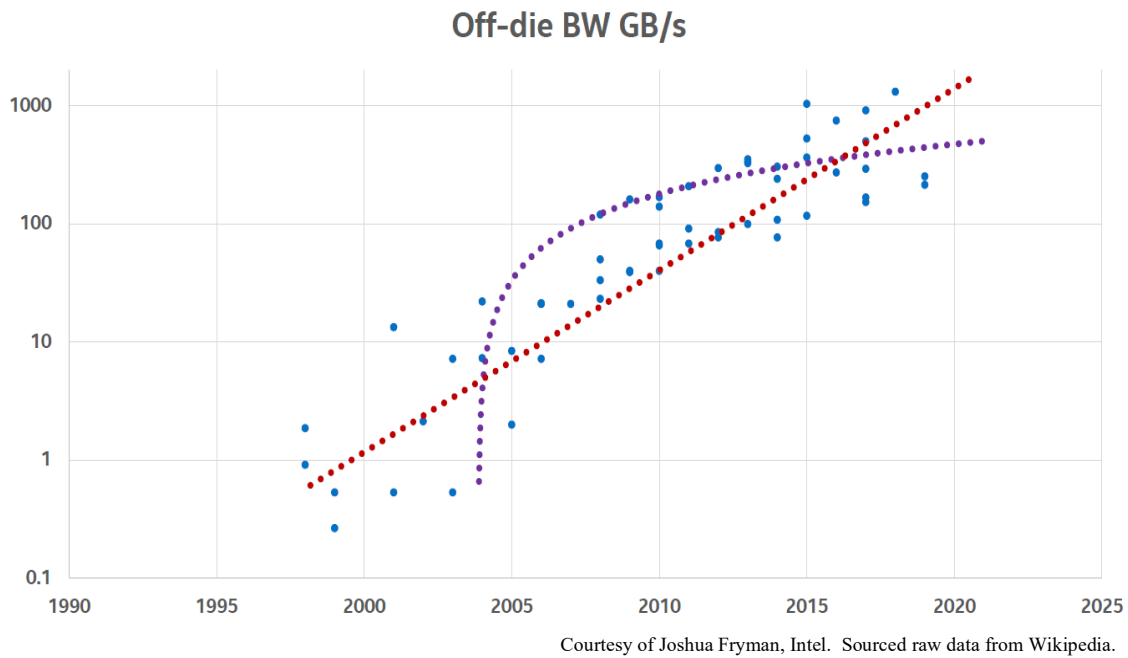

Courtesy of Joshua Fryman, Intel. Sourced raw data from Wikipedia.

**Figure 2**: Off-die bandwidth growth history. Purple dotted line is linear fit, red dotted line is an exponential fit. The red line depicting exponential growth is a better fit to the data.

At the same time, bandwidth demands have been growing exponentially (see Figure 2). Off-die bandwidth already exceeds 1 TB/s, and the bandwidth demands are expected to continue to follow this exponential growth trend. Similar trends are visible in the LAN and device bandwidth requirements. In the longer term, the mismatch between the linear growth of die power and exponential growth of bandwidth demands will lead to limitations on frequency and power.

This exponential growth in off-chip BW has been predicted by Rent's rule, which defines the relationship in an integrated circuit block (whether that block contains a standard cell, functional implementation of logic unit, or the entire periphery of the chip) between the number of external connections T and the number of logic gates N within the box: $T = k\ N^p$. Generally, k and p are constants, with $p < 1$. This mathematical relationship suggests that as the logic count has increased exponentially with time (as anticipated from Moore's Law) I/O demands increase exponentially, as well. While this relationship is not accurate for all computing architectures, it holds for microcontrollers and GPUs (5), likely due to factors such as many-core design, as well as increased memory and network demands to match rapidly increasing instruction execution rates (Amdahl's 2nd law).

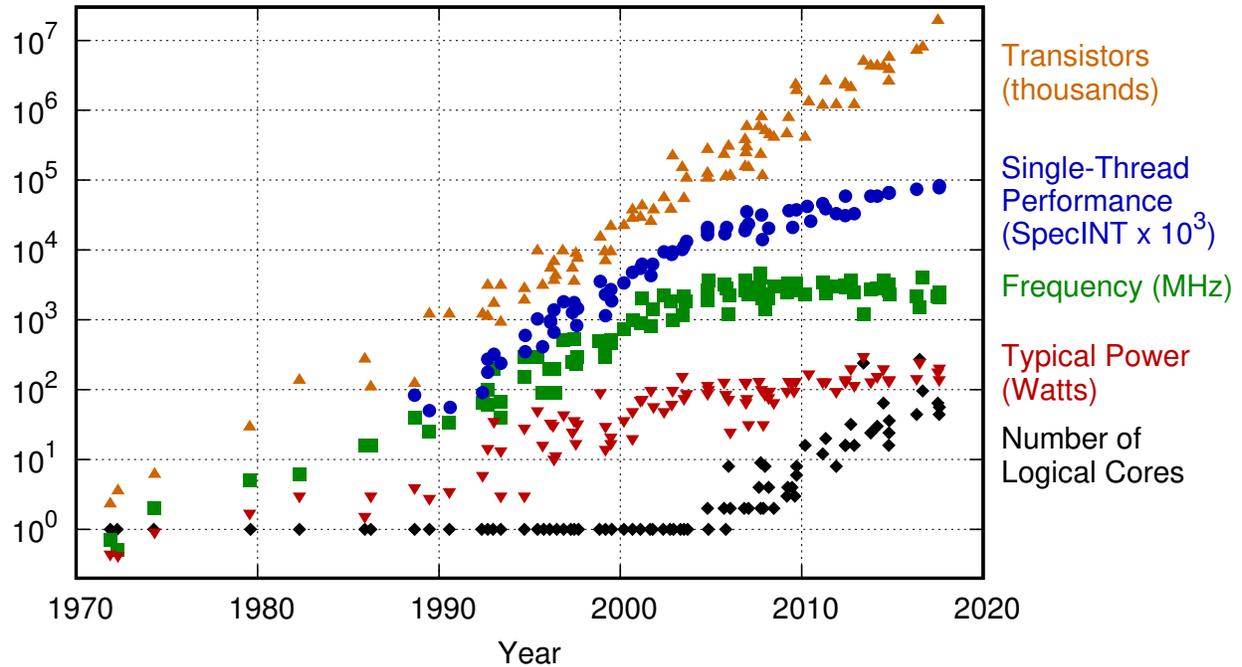

Original data up to the year 2010 collected and plotted by M. Horowitz, F. Labonte, O. Shacham, K. Olukotun, L. Hammond, and C. Batten
New plot and data collected for 2010-2017 by K. Rupp

**Figure 3**: 50 years of microprocessor trend data **(6)**.

Although many predict Moore's law may be slowing or even ending, almost 50 years of data suggests that Moore's law continues to drive transistor counts on chip and has not slowed substantially (Figure 3). There are powerful economic motivations for companies driving Moore's law since exponential performance increases are vital for integrated circuit manufacturers to maintain their competitive edge and commercial markets (7; 8). Therefore, it is likely that exponential growth in transistor logic on-chip (and the resulting increase to instruction execution rates) will continue to drive exponential growth in off-chip I/O.

Maximum interconnect distance is an important parameter that determines the type of technology that is appropriate to use for a given application. Table 2 shows the performance metrics of some example electrical interconnect technologies. In general, longer-reach electrical interconnect technologies are less efficient, with the longest-reach technologies working to around 1-2 meters. The longer-reach technologies also have more latency, largely due to forward error correction (FEC). Although latency can be tolerated in most data center applications, it is not acceptable for high performance computing.

**Table 2:** Example electrical interconnect technologies.

| Metric (2021) | Interposer A | Interposer B | Short/Med Reach MCP | Long Reach MCP | DDR5 @EOL (6.4GT) | HBM2e/3 Interposer | PCIe Gen5 (32GT) |
|---|---|---|---|---|---|---|---|
| Si Density GB/s/mm² | <50 | <150 | <100/<50 | <50 | <8 | <50 | <30 |
| Energy pJ/b | <2 | <0.5 | <1/<2 | <4 | <4 | <2 | <8 |
| Latency 1-way ns | <10 | <5 | <5/<8 | <70 | <10 | <5 | <50 |
| Distance cm | 0.5-7.0 | <0.25 | <3/<7 | <70 | <20 | <1 | <30 |
| Use Case | Si:Si or Pkg:Pkg | Si:Si | Si:Si | Pkg:Pkg | OoPM | IPM | Pkg:Pkg |

Courtesy of Joshua Fryman, Intel. Sourced from various vendors, whitepapers.

For interconnects longer than 1-2 meters, optical interconnects have been the technology of choice. With optics, the performance is essentially independent of distance, but there is a "tax" to convert from the electrical to optical domains that makes it less efficient to use for shorter distances. The origin of this tax is partly related to the architecture surrounding their use. A typical conversion from electrical signals to optical entails going from a serializer/deserializer (SerDes), to peripheral component interconnect bus (PCI), back to SerDes, to the network control center (NCC), back to SerDes, and finally to the optics. This series of conversions both reduces efficiency and increases latency. New technology that reduces this "tax" would make optical interconnects more viable for shorter-reach applications.

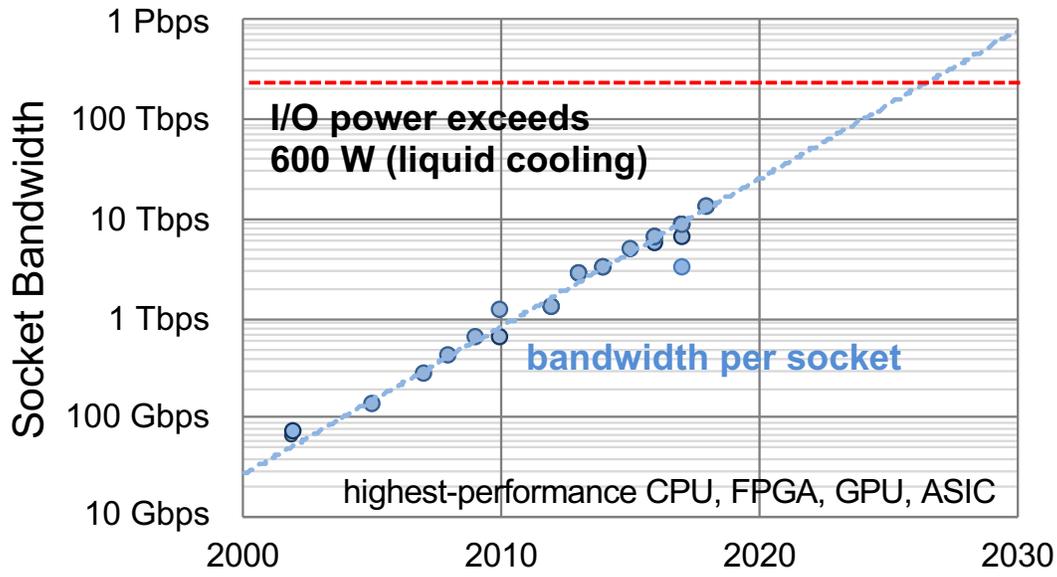

**Figure 4**: Exponential increase in bandwidth requirements for off-chip I/O is quickly exceeding power available per socket (2). Plot assumes a relatively optimist power efficiency of 2.5 pJ/bit (electrical VSR, <10 cm) for off chip I/O.

With the exponential increase in socket bandwidth, we are quickly approaching the point, around 30 Tbps, when the power required to support the I/O requirements will be greater than the total power that is supplied to each socket (2). Higher thermal loads could be tolerated through various engineering solutions, but the cost is prohibitive. Some datacenters are transitioning to water cooling or more exotic cooling strategies, but there is no technology that will be able to handle these exponential issues in the long term. If we extrapolate the current socket bandwidth trend in high-performance CPUs, FPGAs, GPUs, and ASICs, the 2028 single-socket bandwidth target is around 400 Tbps (2), which is beyond the limit achievable with current technology within socket power limits, even with liquid cooling for very short reach, sub 10 cm links (Figure 4). Power requirements to meet socket BW requirements for backplane communication (<7 pJ/bit) in a typical air-cooled configuration will be reached even faster (by 2025). A new technology is needed.

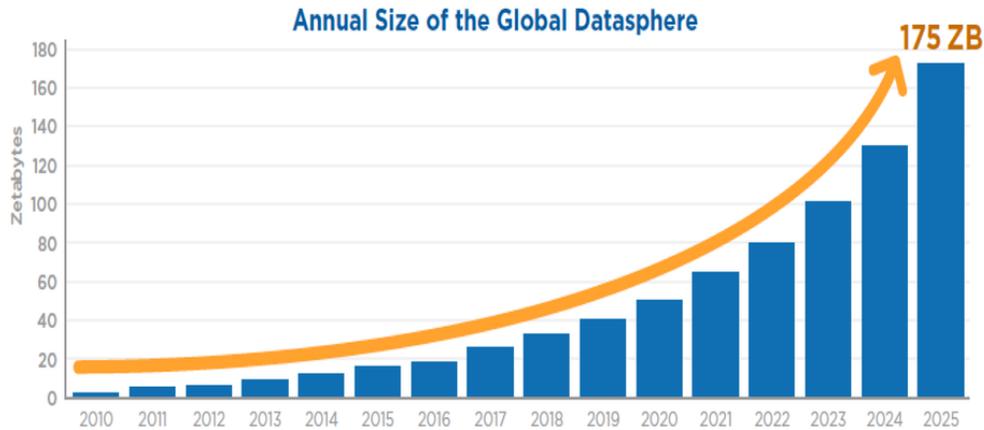
Source: Data Age 2025, sponsored by Seagate with data from IDC Global DataSphere, Nov 2018

**Figure 5:** Predicted size of the global datasphere by 2025 **(9)**.

**Table 3**: Predicted global data usage in 2025.

| Data | ZettaBytes | Bytes | Bits | bps[A] | W[B] | # 400 Gbps |
|---|---|---|---|---|---|---|
| Real-time | 50 | $50 \times 10^{21}$ | $4 \times 10^{24}$ | $1.3 \times 10^{17}$ | $1.3 \times 10^{5}$ | $3.2 \times 10^{5}$ |
| Remainder | 125 | $125 \times 10^{21}$ | $1 \times 10^{25}$ | $3.2 \times 10^{17}$ | $3.2 \times 10^{5}$ | $7.9 \times 10^{5}$ |

Notes:
- A: bps is based on a 365-day year
- B: assuming 1pJ/b technology for point-to-point link with only one connection and one way of communication

| Factor | Inflation Factor | Impact |
|---|---|---|
| Avg. 1-way hop-count | 14x | Raw bps, link count, power, Si area |
| Actual pJ/b per link | 20x | Power |
| O/H: Protocol, ECC, etc. | 2x | All terms |
| Burstiness and Redundancy | 4x | All terms |
| **Total Provisioning** | **~1,000x** | |

Courtesy of Joshua Fryman, Intel.

From the perspective of global data usage, the global datasphere is predicted to be 175 ZettaBytes by 2025 (Figure 5), with about 30% of that being real-time data (9; 10). At first glance, this appears achievable with 400 Gbps links, but it should be noted that these targets do not take into account external factors, such as overhead and redundancy, which together inflate the requirements by around 1000x (see Table 3).

**Table 4:** Socket I/O requirements and factors that may limit them.

| Linear Factor | Exponential Implication |
|---|---|
| Socket Performance | Socket I/O Requirements |
| Signal Density (GB/s/mm$^2$) | Area for I/O Requirements |
| Signal Energy (pJ/b) | Power of I/O |

Datacenter bandwidth demand is on an exponential curve, with a significant fraction of that bandwidth being real-time data. For high-performance computing, real-time data is also a major factor in requirements, which introduces significant technological challenges. These exponential demands will not be met by the linear growth of performance in current technology (see Table 4), and no mature technology is available to address these issues. To meet the 2028 performance targets, we need a solution to be ready in 2021.

## SILICON PHOTONIC SWITCHES

Content provided by Benjamin G. Lee (IBM) and Ming C. Wu (UC Berkeley)

The most advanced high-performance computing (HPC) systems have thousands of compute nodes, and efficiently controlling the data flow connecting these is critical to their operation. Thus, switching technology is yet another area ripe for improvement. Here, we discuss the current electrical switching technology, the possibility of incorporating optical switching, cutting-edge optical switch technologies, and future directions for switching.

Data flow within HPC systems and datacenters is currently controlled with electrical packet switching on application-specific integrated circuits (ASIC). While switch bandwidth has been doubling every 2-3 years, the switching efficiency has been essentially flat since around 2012 (see Figure 6b). Without a major technological breakthrough, switch scaling will hit a thermal wall.

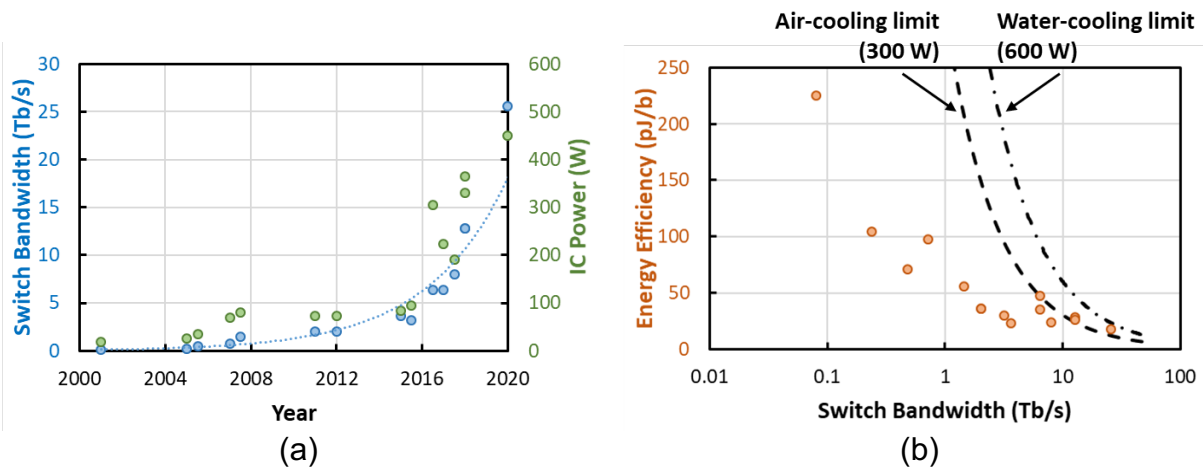

**Figure 6**: (a) Electrical packet switching ASIC historical switching bandwidth. (b) Energy efficiency of switch ASICs with respect to switch bandwidth.

One possible technology that could address this efficiency challenge is optical switching, which has the potential to more seamlessly connect optical interconnects with first-level packages (see Figure 7). Today, the electrical switch ASIC is packaged onto a first-level package, with > 10 cm transmission lines stretching to on-board or edge-of-card optics. Here, bandwidth is limited by the pin density of the board and efficiency by the length of the transmission lines. As technology moves toward co-packaged optics, where optical transceivers are integrated with the switch ASIC on a first-level package, the switch ASIC signal will travel over ~1 cm transmission lines and bandwidth will be limited by the pin density of the package. Implementing optical switching would eliminate the need for a high-bandwidth electrical switching ASIC, removing the energy-intensive high-speed transmission lines and the bandwidth-limiting pin constraints altogether; instead, the optical switch would transparently route the high-bandwidth optical interconnects via low-bandwidth electronic controls interfacing to the optical switch through the first-level package. With this innovation, the bandwidth would be limited by the (large) optical spectrum, and the switch energy would be expended steering pipes of data instead of interacting with bits. As a result, such a switch would be agnostic to data rates and formats. For this to be a feasible technology, both the associated latency and cost must be competitive with the current technology.

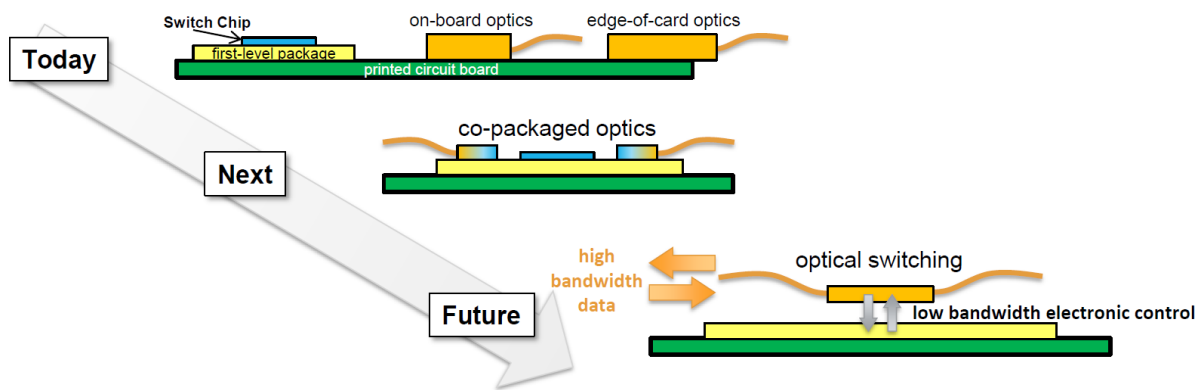

**Figure 7**: Switching technology trajectory.

One of the obstacles to moving toward optical switches is that an optical switch would be a circuit switch, not a packet switch. This introduces new challenges, including latency, scheduling and phase locking. In short, optical circuit switching (OCS) is not a simple drop-in replacement for electrical packet switching. More investigation is needed to determine which use cases and what optimization spaces are most suited for OCS solutions. For example, leaf switches (level 2) need large port counts and can get by with slower speeds and somewhat higher cost, whereas top of rack switches (level 1) need fast reconfigurability, low power, and very low cost while requiring few ports. Other areas, such as communication to storage and the network, will have their own optimization parameters. The use of circuit-switched networking would represent a major technology shift in HPC system architecture and a major design challenge for HPC network development and potentially for application development, as well. Application data flows will almost certainly be seriously impacted by the use of circuit-switched as opposed packet-switched networks. In addition, new device and system level photonic simulation tools will be needed to develop network topologies, routing, and controls for circuit-switched, stateless (no memory buffers in the switch) routing in HPC systems.

For all of these operational conditions, it is important to understand the critical parameters for an optical switch. In general, optical switching needs to be fast compared to the data transfer size, as the switching speed places a limit on the throughput and latency. Figure 8 shows how different switching times (due to configuration, control, and scheduling) affect throughput at different data rates. For example, a switching time of 10 ms achieves only 50% throughput for a data transfer size of 1 GB at 400 Gbps. As a result, this switching speed would only be appropriate for long-lived flows, which would not often need to be reconfigured. It is also important that an optical switch be cost-efficient in order to be an attractive alternative to electrical packet switching. Optical switching technologies can reduce cost by leveraging commercial photonic foundries and high-throughput microelectronic assembly tooling. These will be especially important because of the large number of analog electrical connections and fiber attachments.

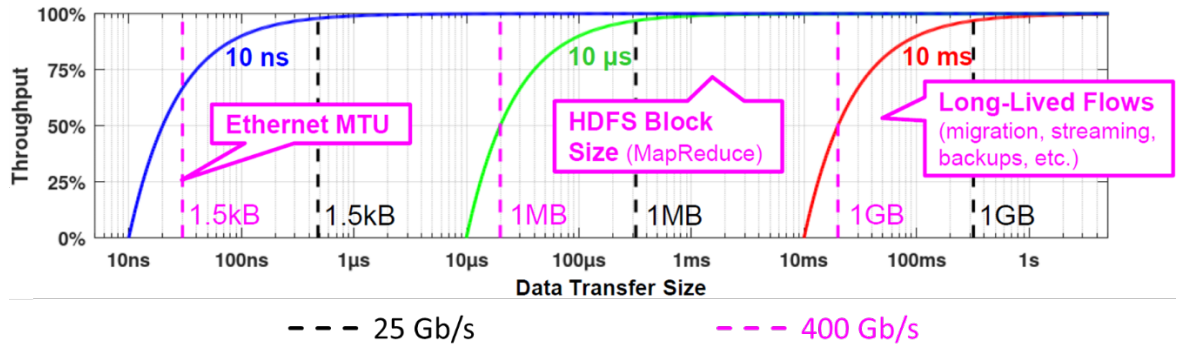

**Figure 8**: An illustration of the effect of switching speed on throughput for different data transfer sizes.

The most mature optical switching technologies are free-space or silica-based and are commercially available. While they feature high extinction ratios, polarization insensitivity, hundreds of ports, and just a few dBs of insertion loss, their switching speed is on the order of 1 ms or greater. In the research domain, new integrated-photonics-based switch technologies show potential to achieve much faster switching speeds with lower power and lower cost, but there are a number of performance hurdles to be overcome, including port count, insertion loss, crosstalk, and polarization handling.

**Table 5:** Best demonstrated silicon 2×2 switch cell performances.

|  | Thermo-Optic MZI | Electro-Optic MZI | Thermo-Optic Ring Resonator | Photonic MEMS |
|---|---|---|---|---|
| Loss | < 0.15 dB | 0.8 dB | 1.1 dB (drop) | 0.026 dB (through), 0.47 dB (drop) |
| Transient | > 10 µs | 4 ns | 25 µs | 0.9 µs |
| Extinction | -30 dB to -40 dB (-50 dB with tunable couplers) | -28 dB (-38 dB with nested) | -20 dB (through), -40 dB (drop) | -60 dB |
| References | (11; 12; 13) | (14; 15; 16) | (17) | (18) |

As we examine these integrated photonic technologies, it is helpful to compare them based on the 2x2 switch cell, which is the fundamental building block of a switching network. Photonic switching technologies that are being explored include thermo-optic Mach-Zehnder interferometers (MZIs), electro-optic MZIs, thermo-optic ring resonators, and photonic micro-electromechanical systems (MEMS) switches. Table 5 shows some demonstrated performance metrics for each of these. In addition to the switch cell technology, switch topology plays an important role in switch performance. Table 6 compares three switching topologies. The appropriate switch topology somewhat depends on the strengths and weakness of the 2x2 switch cell in use.

**Table 6:** A comparison of strictly non-blocking (SNB) switching topologies: crosspoint switch matrix (CSM), path-independent insertion loss (PILOSS), and double-layer network (DLN) **(19)**.

|  | Total Switches | Switches in Path | Crossings in Path | $1^{st}$-Order Cross-Talk | Best Use |
|---|---|---|---|---|---|
| **CSM** | $N^2$ | 2N-1 | 0 | N-1 | Cells with asymmetric loss |
| **PI-LOSS** | $N^2$ | N | N-1 | N-2 | Tight loss distribution on outputs |
| **DLN** | $5/4\ N^2-2N$ | $2\log_2(N)-1$ | $3N-2\log_2(N)-4$ | 1 | Minimizing crosstalk; Beneš-like scaling |

The highest switching-speed photonic technology to date has been the electro-optic MZI, with switching times on the order of 4 ns. This switch element has been scaled to a fully packaged 8x8 switch module based on a double-layer switching topology, achieving fiber-to-fiber loss between 7.5 and 10.5 dB and crosstalk below 30 dB while consuming 1.9 W including drivers (20).

Realistically, scaling to much higher port counts requires extremely low insertion loss per switching cell, and photonic MEMS switches are a promising candidate for addressing this need. The crosstalk, bandwidth, and loss in MEMS switches is potentially better than other technologies, and Figure 9 shows how the MEMS switch low loss contributes to better scalability. MEMS photonic switches have been demonstrated up to a 240x240 port count with a total insertion loss of 9.8 dB, 400 ns switching time, and crosstalk < -60 dB (21). More development is needed, however, to further reduce the insertion loss and increase the port count to practical levels.

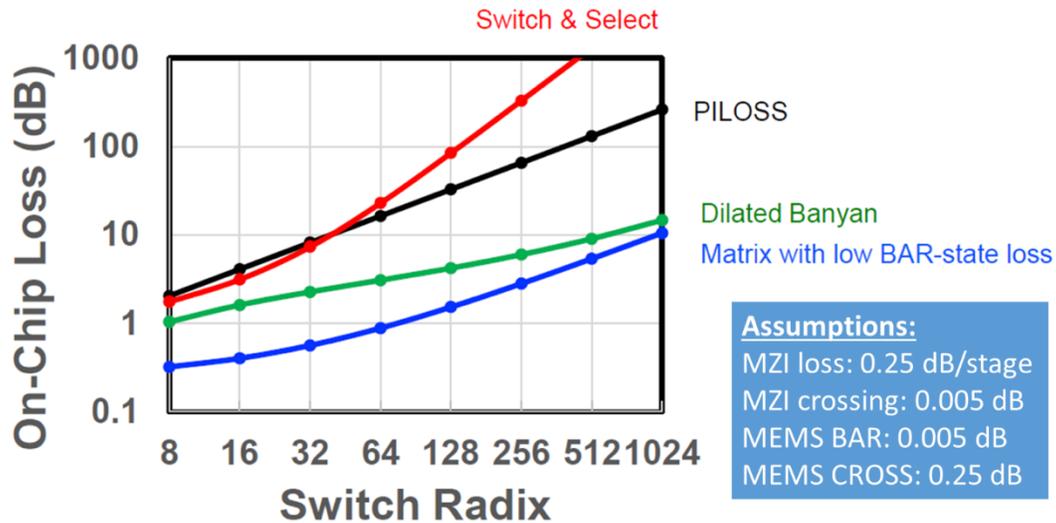

**Figure 9**: Scaling of switch loss for various topologies. Matrix loss trend assumes MEMS 2x2 switch cell, all others assume MZI 2x2 switch cell.

Beyond scaling port count, there are a number of future directions photonic switching research could take to expand performance and application space. The current major gaps in performance for photonic switching technologies are polarization handling and loss. There have been a few attempts to decrease polarization dependence, including in MEMS (22) and thermo-optic MZIs (23), but in general, polarization-dependent loss is still too high. The total insertion loss is also prohibitive for practical applications. In addition to better optical design to reduce this loss, it may be beneficial to integrate gain into the switch assembly to compensate for this loss (24; 25), but the tradeoff with manufacturability and reliability challenges should also be considered. Insertion loss is also strongly dependent on fiber-to-chip coupling loss, which becomes an even bigger challenge as port counts increase. Much improvement is needed in automated optical packaging to enable low-cost, ultra-efficient, high-port-count multi-fiber coupling.

Switching is an important component of links in HPC systems, but the current electrical paradigm is approaching thermal limits. New technologies are needed, and optical switching is a potential path beyond the thermal wall. Moving to optical switching will require some architectural changes, though, since it is most suited to circuit switching instead of packet switching. Optical switching technologies are in development to address some of the challenges in these systems, but more development is needed, especially to reduce insertion loss, increase speed, and reduce cost to make these feasible improvements compared to electrical packet switching ASICs.

# STATE-OF-THE-ART ELECTRICAL INTERCONNECTS

Content provided by Jonathan Proesel (IBM) and Mounir Meghelli (IBM)

As the performance demands on interconnects increase, electrical interconnect technology roadmaps are rising to the demand, targeting higher data rates. Electrical interconnects are attractive due to their very low cost, on the order of cents per Gb/s in the drawer and tens of cents per Gb/s at the direct attach level, and this cost per b/s continues to go down as data rate increases.

**Table 7**: The common electrical I/O (CEI) 56G standard as an example of the reach/modulation/power tradeoff **(26)**.

| Standard | Application | Reach | Mod. | FEC | Efficiency |
|---|---|---|---|---|---|
| USR | 2.5D/3D | 1 cm | NRZ | none | < 1 pJ/b |
| XSR | Chip to nearby optics engine | 5 cm | NRZ PAM4 | light | < 1.5 pJ/b |
| VSR | Chip to module | 10 cm | PAM4 NRZ | yes | < 2.5 pJ/b |
| MR | Chip to chip and midrange backplane | 50 cm | PAM4 NRZ? | yes | < 5 pJ/b |
| LR | Interface for chip to chip over backplane | 100 cm | PAM4 ENRZ | strong | < 7 pJ/b |

However, as data rate increases, electrical link technologies encounter tradeoffs between reach, modulation, and power. A good example of this is the Common Electrical I/O (CEI) 56G standard established in the Optical Internetworking Forum (OIF) Next Generation Interconnect Framework (Table 7). Here we see that the energy per bit goes from <1 pJ/bit for ultra-short reach (USR) to <7 pJ/bit with advanced modulation and strong forward error correction (FEC) for long reach (LR). Because of the varying needs for FEC at each distance, the latency in each of these examples is different. In datacenter applications, increased latency has become more accepted (although still an important consideration for certain key applications such as providing search results), and architectures are being developed to work around it. In contrast, high performance computing applications are inherently intolerant of high latency, so some of these electrical interfaces may not be workable.

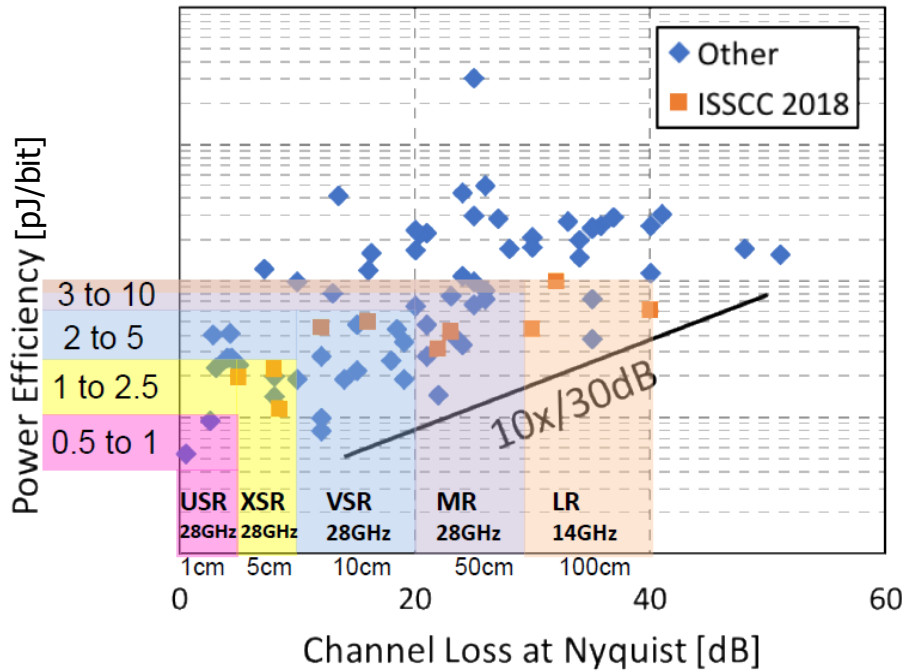

Figure 10: Electrical wireline research: power, loss, and distance (27).

The next goal for the industry is 100G signaling, the CEI standards for which are still under development. Looking at the electrical wireline research trends (27), we see that data rate has been increasing with smaller process nodes, as expected. However, the links are not becoming more efficient. Mapping the CEI-56G standards onto the efficiency vs. loss graph (Figure 10), we see that we are already at the edge of the efficiency that has been demonstrated.

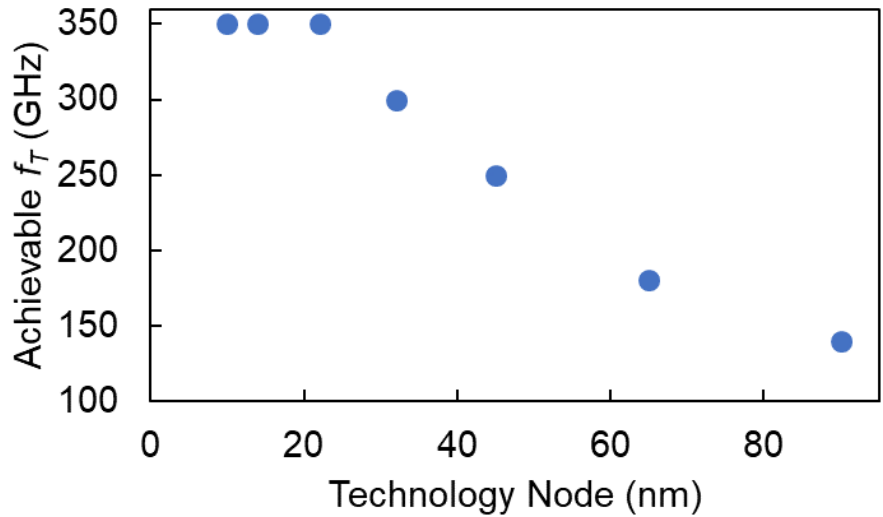

**Figure 11**: Analog performance of CMOS with respect to process node (courtesy of J. Proesel and M. Meghelli, IBM).

In order to reach 100G signaling in electrical links, there are a number of challenges that need to be addressed. CMOS technology is not scaling in the same way it has been, with longer time between nodes and analog performance not getting much faster with each node (see Figure 11). Channel capacity and packaging technologies also present challenges.

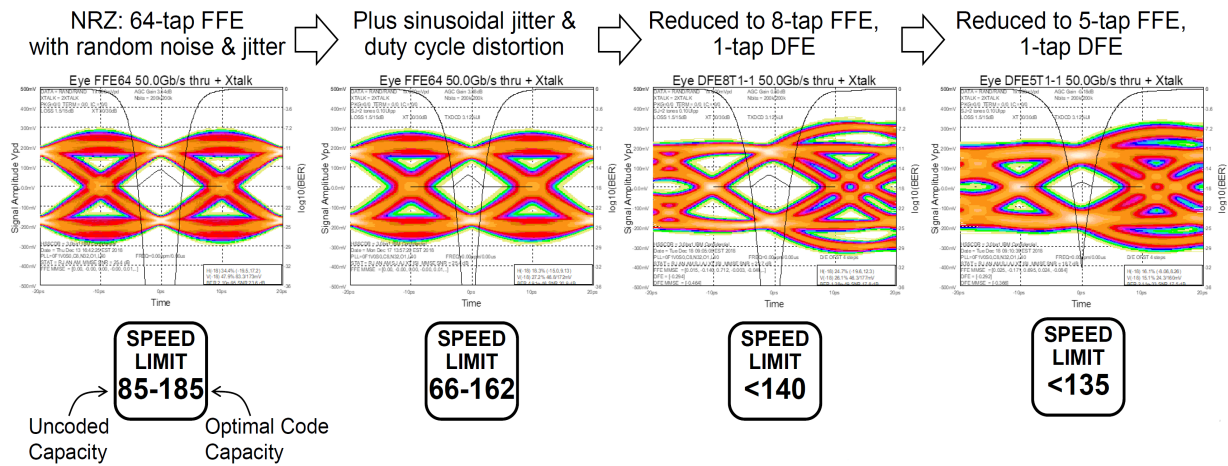

**Figure 12**: Simulation of a medium reach (MR) link with 25dB loss at 25GHz as nonidealities are added. (courtesy of Troy Beukema, IBM).

The maximum channel data rate is defined in information theory as the Shannon capacity. For existing LR electrical channels, taking into account channel loss and assuming additive white Gaussian noise (AWGN), the Shannon capacity is 200 Gbps. This theoretical value does not take into account other nonidealities in the link, so a real-world system requires additional margin beyond the Shannon capacity. As an example, we simulate a medium reach (MR) link having 25 dB loss at 25GHz with a variety of nonidealities added, shown in Figure 12. When we add random noise, jitter, duty cycle distortion, five-tap feed-forward equalization (FFE), and single-tap decision feedback equalization (DFE), the actual coded capacity is reduced to 135 Gbps (uncoded capacity cannot be calculated). For LR links, the available margin is rapidly approaching zero, as shown in Figure 13. The exact required margin for a real-world link is up for debate; regardless, reaching 112 Gbps will be a major challenge. New technologies can help address this, including advanced modulation formats and better link materials.

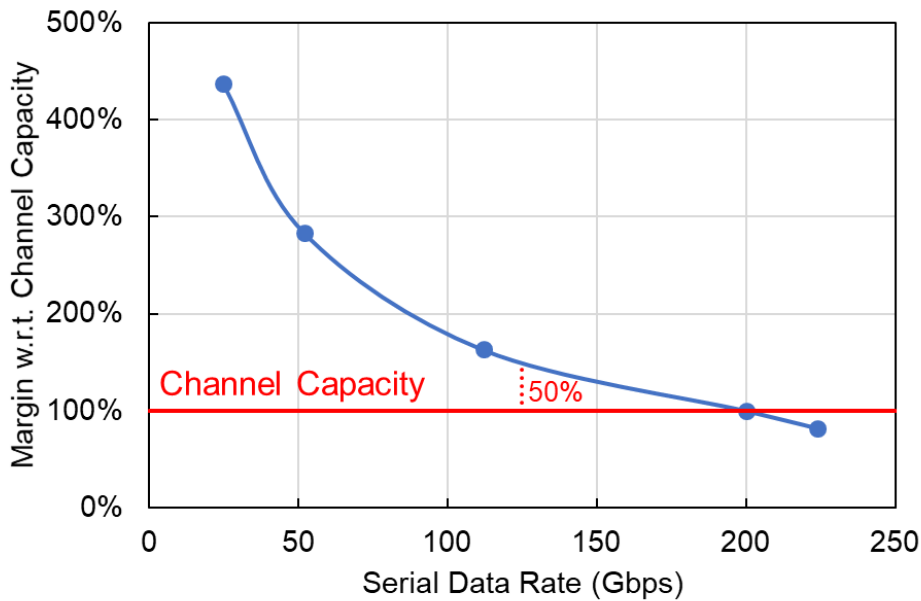

**Figure 13**: Margin beyond Shannon capacity with respect to data rate for a LR channel.

One example of an advanced modulation format that is being explored as links approach the Shannon capacity is pulse-amplitude modulation (PAM), in which data is encoded in more than two voltage levels (4, 8, or 16). This would allow lower loss for a given data rate because the equivalent baud rate for PAM is lower. Another possibility is discrete multi-tone (DMT) modulation, which can increase bandwidth by using an RF-like approach. Employing more advanced FEC algorithms, such as Reed-Solomon or Bose-Chaudhuri-Hocquenghem, can also help extend reach of an electrical link. In all of these cases, the gains conferred by these technologies come at a price: increased power, increased chip area, and increased latency (from additional en/decoding and FEC). As data rates scale, electrical links will have to either be shorter or use better materials to avoid increasing power and chip area costs.

One material-related innovation that is starting to be adopted by industry is the use of twin-axial flyover cables to replace printed circuit board (PCB) connections (28; 29). Because these cables are made of better materials than PCB, frequency-dependent loss is reduced, and their reach is up to 2 m. Using flyover cables does limit the edge density of links due to the physical size of the twin-axial cable, and the cost is higher than traditional PCB interconnects. There may be additional hidden costs because each cable connector will stress the board, and reengineering the board to accommodate this stress will likely be necessary. While flyover cables are a promising technology for improving electrical link performance through better materials, they have not yet been publicly demonstrated as a 100G link. Improving the PCB material itself is also a possibility, but PCBs are already approaching the intrinsic material limits of dielectric quality and copper roughness. Improvements in this area will result in small performance gains.

Chip I/O bandwidth density is also limited by the pinout. Architecture is at least one contributor to this limitation. For example, with standard differential signaling, one signal requires four to six pins. New architectures, such as single-ended (e.g. NVIDIA ground-referenced signaling) and multi-wire encoding (e.g. Kandou Bus Ensemble NRZ), can increase bandwidth density, but these changes will only approximately double density once; they will not address the continued I/O density scaling in the long term. Additional scaling is possible with decreased pad pitch and 3D integration. 3D integration, in particular, is an area of active research (30; 31; 32), but it is important to note that, as the number of integrated layers increases, yield decreases and cost increases. Eventually, the cost of integrating an additional layer will be too high to justify the marginal increase in I/O density.

As a whole, the scaling of data rates in electrical links is limited by their reach. At shorter distances, this scaling can continue through innovations in advanced modulation, FEC, packaging, materials, and architecture. At these shorter distances, the cost and efficiency of electrical links is better than optical links. However, the link distance over which these technologies will be applicable is short and getting shorter with increasing data rates. For long reach (1 m) links, we are already approaching the theoretical Shannon limit, and all of these proposed technological improvements are marginal at best. The question is when/at what distance this limit will be reached, at which point switching to optical links will become cost-effective.

# STATE OF THE ART OF PHOTONICALLY-ENABLED SOCKET-LEVEL I/O

Content provided by Tony Lentine (Sandia) and Marc Taubenblatt (IBM)

Optical interconnects are currently being used extensively between racks in datacenters, where the reach is on the order of 20 m up to 2 km (see Figure 14). The next generation of optical links are moving to within the rack, e.g. server to the top of the rack switch (TOR), but the comparative cost is a significant limitation, given the, to date, much lower cost of electrical links. Here, we discuss some of the factors contributing to the relatively high cost of optics and the tradeoffs involved in the various optical interconnect technologies.

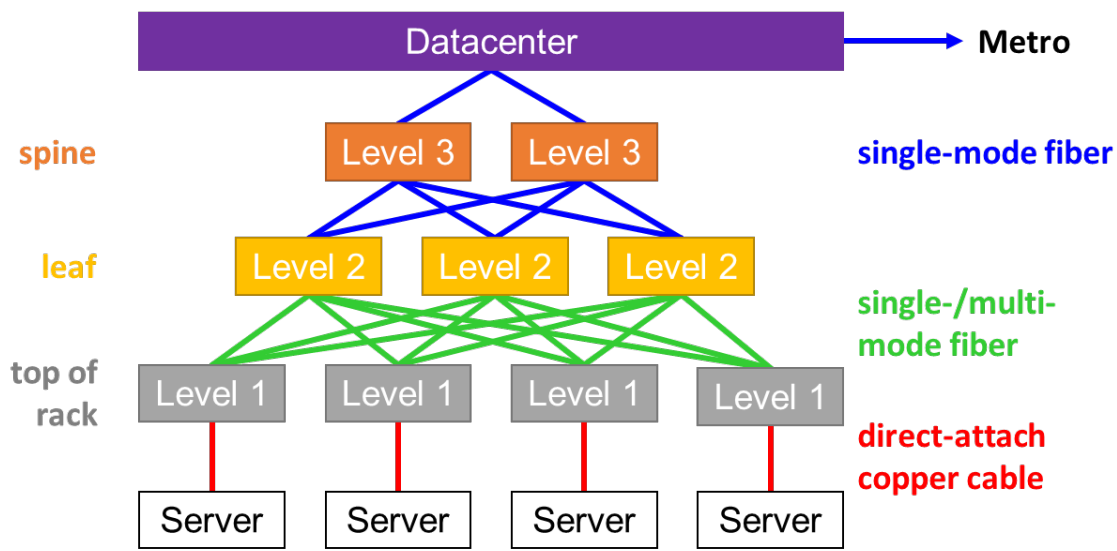

**Figure 14:** Typical links in datacenter **(33)**.

The optical interconnect market is being driven by the demands of large-scale cloud data centers due to their large market volume, which is focused on links from 10's of m up to 2 km and some metropolitan area networks (80 km links and beyond). As single-chip bandwidth capabilities continue to increase, the demand for optical interconnects within the rack environment will likely increase (e.g. server to top-of-rack switch (TOR) connectivity. Unlike datacenter interconnect applications, HPC primarily utilizes optical links shorter than 50 m. However, the HPC market volume is much smaller, and thus does not have as much leverage to drive optical interconnect vendor offerings. The HPC market, so far, has continued to make use of the more cost effective, though shorter reach VCSEL-based optical interconnects, which have long been the mainstay of traditional data centers (and are today still used significantly for shorter links in cloud data centers as well).

**Table 8:** Details on each level of link in the datacenter system **(34; 35)**.

| Link | Reach | Volume | Medium | Cost/Gbps |
|---|---|---|---|---|
| CR to Metro | 10-80 km | 100s | SMF | $10-100s |
| Spine to DCR | 1 km | 100s | SMF | $1-5 |
| Leaf to Spine | 400 m | 1000s | SMF | $1s |
| TOR to Leaf | 20 m | 1000s | SMF/MMF | $1s |
| Server to TOR | 3 m | 10000s | Copper | $0.10s |
| Intra-Server (PCB) | < 1 m | 1000000s | Copper | $0.01s |

For all of these applications, cost continues to be the main driver of adoption, and thus when lower-cost copper interconnects are sufficient, this has been the technology of choice. There are many factors contributing to the high cost of optics, including cost of materials, heterogenous assembly, test, yield, and reliability. One of the reasons these costs are high is that there are too many variations in components, so the high volume of demand is divided among many custom solutions. That said, optical costs continue to drop and as it becomes more and more difficult for copper to meet bandwidth, speed, and distance requirements (e.g. 3m), optics will begin to make inroads. Having more well-defined standards may help vendors better take advantage of the large market and drive costs down, though not without the added time burden of standards development.

In addition to lower cost, the market is driving toward higher data rates. The roadmap for ethernet link speeds (36) calls for over 1 Tbps links around 2025. Achieving this aggregate bandwidth is a technological challenge, which will be addressed via a number of different strategies (see Figure 15). Higher per-lane data rate using NRZ signaling has historically been the lowest-cost way to increase aggregate bandwidth. However, as desired per-channel data rates are beginning to exceed 100 Gbps, it has become difficult to maintain this trend. More feasible strategies for these higher bandwidths include advanced modulation, such as pulse amplitude modulation (PAM) and coherent communication, or parallelization, such as using multiple fibers or wavelengths. It should be noted that each of these strategies will require different degrees of forward error correction (FEC), which contributes to increases in latency. To date, Ethernet applications in data centers have been more tolerant of the additional (e.g. 100ns) latencies introduced by the FEC required for PAM4, as these are small compared to the many microseconds of latency across the full software stack. However, many HPC and specialized computing applications will be far less tolerant of this additional latency.

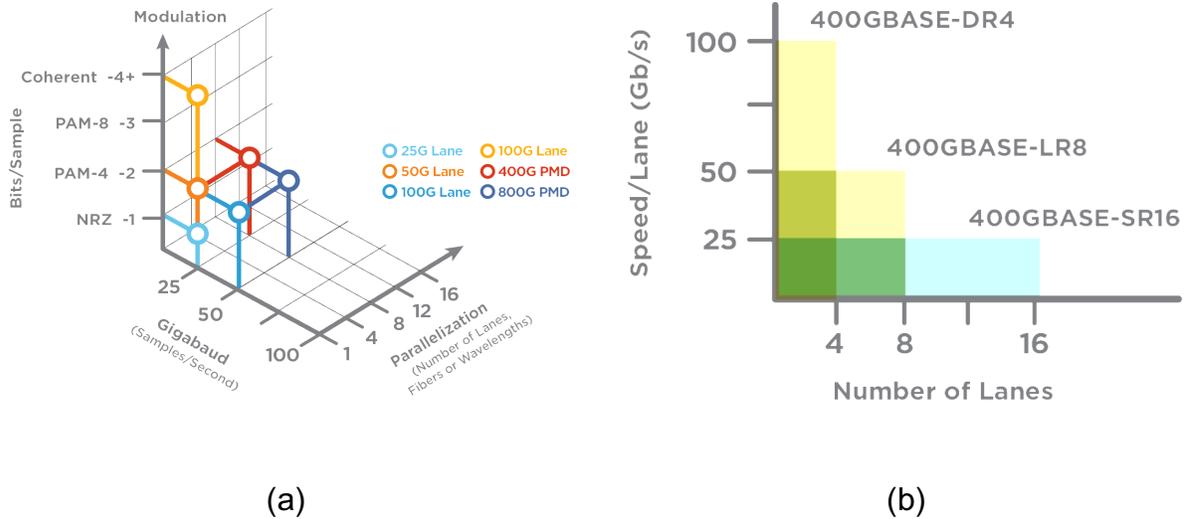

(a)          (b)

**Figure 15:** (a) Higher aggregate bandwidth can be achieved by increasing sampling rate or number of lanes, or by choosing a different modulation technique. (b) A given ling speed can be achieved by choosing different combinations of number of lanes and speed/lane **(36)**.

Currently-used optical link technologies fall into two categories based on fiber type: multi-mode fiber (MMF) and single-mode fiber (SMF). In MMF, the light source technology is typically directly-modulated vertical cavity surface emitting lasers (VCSELs) with wavelength around 850 nm. Parallelization is accomplished through parallel fibers or thru a coarse wavelength-division multiplexing (e.g. shortwave wavelength division multiplexing is one such multisource agreement standard, supporting four wavelengths). This transceiver ecosystem has historically been widely used in datacenters with lower cost and power due to factors such as the drive simplicity and high alignment tolerance. But as channel baud rates have steadily increased, the distance capability of MMF has shortened from what was once >300m at 10Gbps to less than ~50m at 56Gbps. At the same time, data center size has increased, and multi-building campuses are common, requiring distances of up to 2 km. In general, although reliability is an issue for all semiconductor lasers, VCSELs are particularly sensitive to elevated temperatures due to their typically high current density, which exacerbates defect formation. In addition to maintaining case temperatures (typically < 70C), adding spare channels or spare VCSELs is another option to address shortened lifetimes if elevated temperature operation is required.

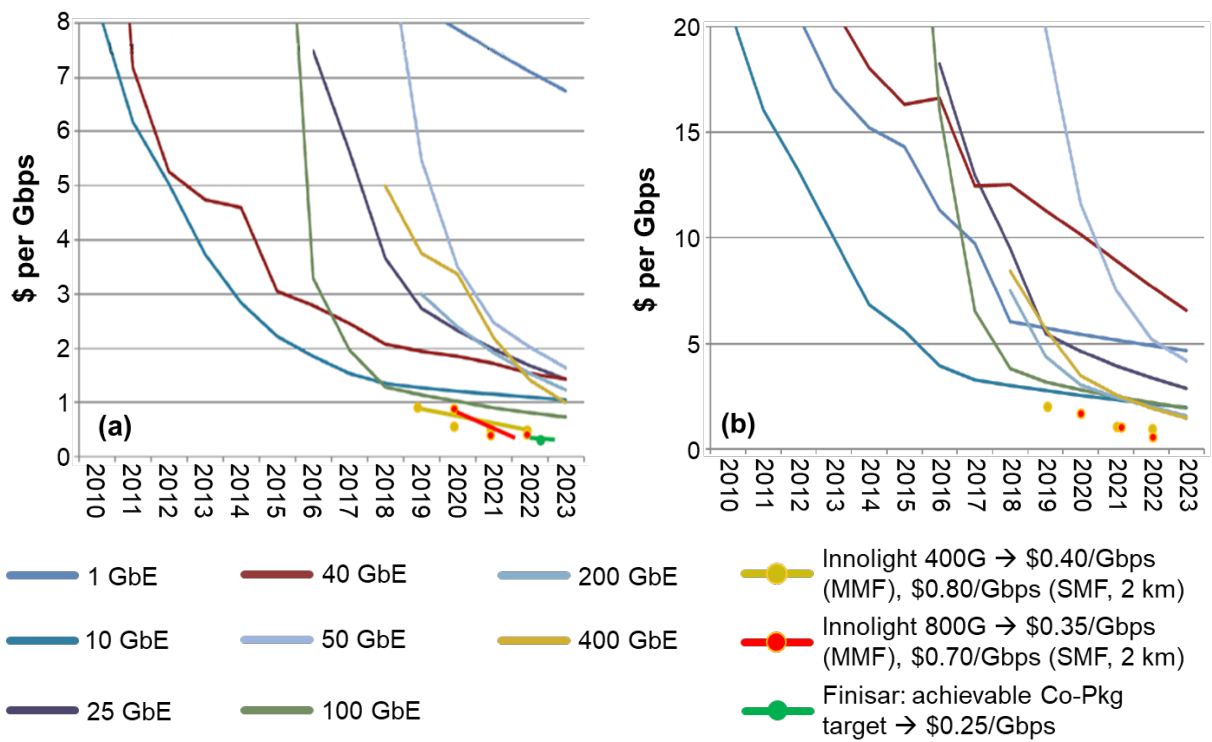

**Figure 16**: Cost/Gbps for ethernet, all form factors included. Graphs presented publicly at **(37)**, with data from refs. **(38; 39)** (a) Ethernet short reach (MMF) by data rate. (b) Ethernet 0.5-10 km reach by data rate.

SMF-based optical transceivers have traditionally addressed the telecom market where very long distances and expensive components can be tolerated. More recently, silicon photonics based SMF transceivers have emerged as a modest cost solution addressing the up to 2 km distance range (as well as longer distances such as 80km coherent links). SMF enables longer reach than MMF, and single-mode photodetectors, due to their smaller size, have lower parasitics than multimode photodetectors. The most common platforms for silicon photonics based SMF transceivers utilize Mach-Zehnder or electro-absorption modulation schemes instead of directly-modulated lasers. More recently, there have been announcements for more compact and power-efficient resonant ring-based modulators as (40; 41). Si photonics can take advantage of integrated-circuit-like mass production, and can support dense wavelength-division multiplexing (DWDM). However, to date it has been more cost-efficient to stay with coarse wavelength spacing at 4 wavelengths, and coherent modulation schemes, despite costs that are higher than in MMF systems. A major pain point in the Si photonics/SMF system is packaging, where the tighter tolerances required for SMF require more expensive packaging techniques. Another pain point is low power efficiency, which is due to a combination of lower efficiency laser sources, fiber-to-chip coupling and on-chip losses. (The disparity in efficiency between MMF and SMF systems has been shrinking, however.) In addition, Si photonic devices are usually polarization-dependent, necessitating additional on-chip polarization handling components (with some loss) to enable standard non-polarized fiber to be used for links. These costs and limitations have made SMF less appealing than MMF for many cost-sensitive applications, but increasing data rates may change this.

Although optical links are expensive, the cost with respect to bandwidth has been dropping over time (see Figure 16), to between about $1/Gbps and $5/Gbps, depending on data rate, for ethernet short reach (MMF) in 2019, and between about $1/Gbps and $13/Gbps, depending on data rate, for ethernet 0.5-10 km reach (38; 35). For 2022, Innolight is aiming for $0.40/Gbps at 400G and $0.35/Gbps at 800G for ethernet short reach, and $0.80/Gbps at 400G and $0.70/Gbps at 800G at 2 km reach (39). Similarly, Finisar is aiming for $0.25/Gbps in a co-packaged form factor by 2022 (37; 39). The current cost could soon be competitive with active electrical cables, but passive electrical cables are still the lowest cost. Due to cost, electrical links are preferred if they can meet the performance demands, but they are not able to meet link requirements for higher data rates and longer distances. Figure 17 shows some representative optical and electrical links at different data rates and reaches, with the regimes most applicable to HPC, datacenters, and metropolitan area networks highlighted. To a certain extent, there are too many variables to give a hard boundary for switching to optical links, but the general rule is for longer reach and higher data rates.

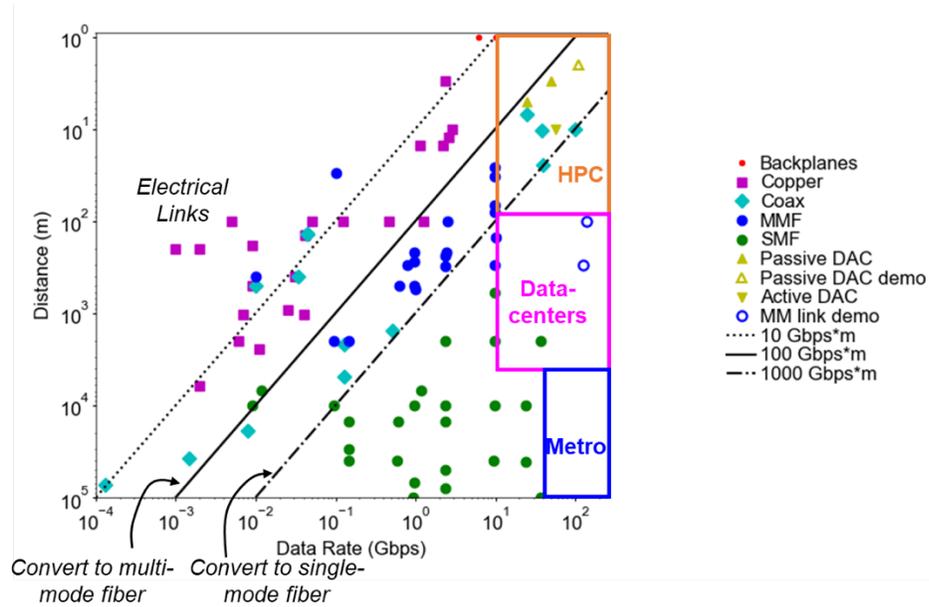

**Figure 17**: Comparison of various links' performance at different data rates and reaches **(42; 43; 44; 45; 46; 47)**.

Packaging of optical links is one area that is being actively explored for increasing data rates and reducing costs of interconnects. A major question is physically where an optical module should be with respect to electrical components. Figure 18 illustrates the main options, each of which presents some tradeoffs related to data rates, replaceability, and thermal controls.

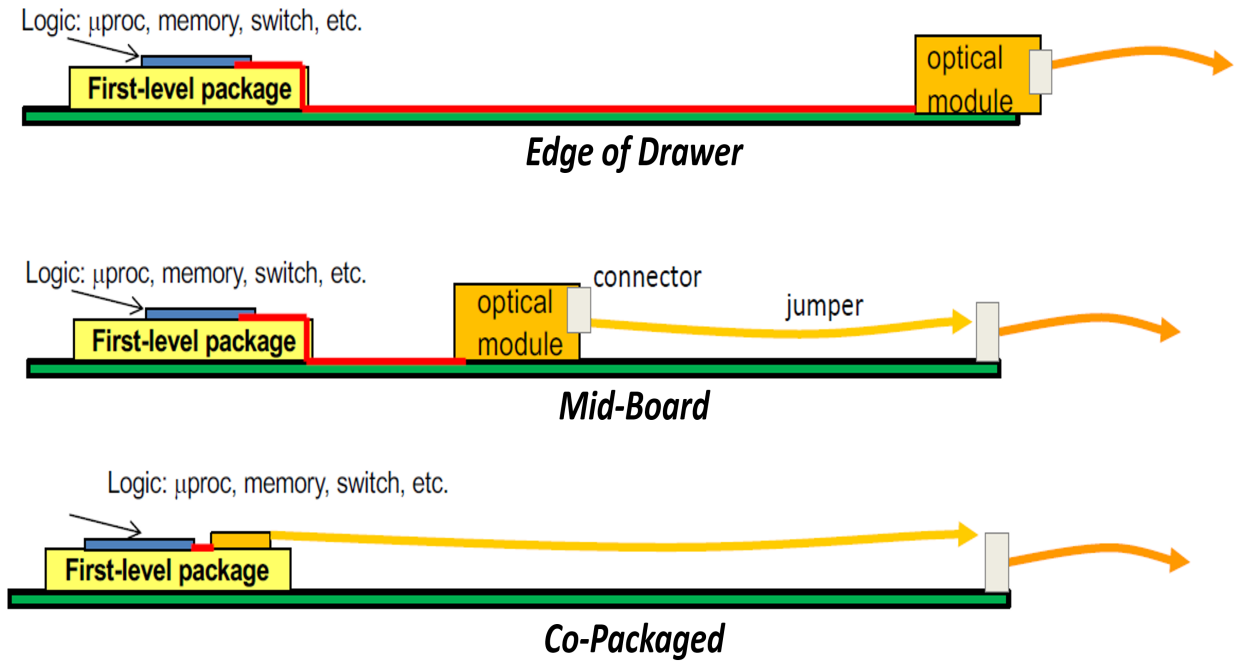

**Figure 18**: Optics packaging options.

Placing the optical components at the edge of the drawer makes them easier to replace in the event of failure, but the electrical link must be long, which is especially problematic at higher channel data rates. Moving to the mid-board shortens the electrical link but makes the optical module less accessible/replaceable. Standardization of this strategy is being pursued by the Consortium for On-Board Optics (COBO).

Co-packaging the optics with the first-level package is more promising for lower power and higher data rates. However, if not carefully co-designed, it could increase design costs, reduce yield, interfere with cooling, and eliminate the ability to independently replace the optical components. There is some industry interest in this strategy; Microsoft and Facebook have launched the Co-Packaged Optics (CPO) Collaboration, which is developing standards for this arrangement. Some examples of implementations include IBM/Finisar's MOTION, which uses VCSELs and MMF (48), and Xilinx's on-board optics solution (49) and Rockley Photonics's OptoASIC (50), both of which employ Si photonics. One potential hidden cost associated with these implementations is that the optical connections come out of all sides of the assembly, making in-drawer routing a challenge. Heat sink size should also be taken into account.

2.5D/3D integration is a much less developed technique that would further shrink the electrical links, but it could also further increase costs due to design, yield, cooling, and replaceability. Yet, it may be the only approach that enables the continued growth in capacity of data center switching chips with aggregate I/O bandwidths above 50 – 100 Tb/s.

Fiber-to-chip coupling is another packaging-related challenge, which is more specific to Si photonics. The optical mode size in a Si waveguide is an order of magnitude smaller than the mode size in an optical fiber, so a low-loss method for converting between the modes is required. In addition, a scalable solution must couple multiple fibers to each chip using an automated assembly process. Strategies for this coupling can be broken down into three categories: edge coupling, grating coupling, and adiabatic coupling, each of which has been implemented in industry. Edge coupling is broadband, polarization-independent, and has shown losses as low as 0.7 dB for a single fiber, but the alignment tolerance is very low, so automated assembly will result is degraded performance (1-3 dB) (51). Grating couplers are easier to align and require less chip processing than edge couplers, and they can be designed to effectively handle polarization diversity (52). However, they are wavelength-dependent and exhibit higher coupling loss. Adiabatic coupling is less well-developed, but shows similar performance to edge couplers with less chip processing (53). None of these has emerged as the preferred option because they all have high costs for assembly, requiring specialized equipment, and even the lowest-loss couplers have > 1 dB loss per facet when assembled in a scalable manner.

In conclusion, optical links are already being employed extensively in data centers, especially for the longer links between racks, and the growth of these areas is being driven by cloud-related applications. Because of the lower cost of electrical links, they are generally preferred over optics if they can meet the performance needs, but as data rates increase, the need for optics at relatively short distances < 1 meter will increase as well. Some of the dominant cost drivers in optical components are transceiver power, module size/packaging, assembly, and physical connections. Standardizing form factors and increasing reliability may help reduce these costs, but more work is needed to make optics an economically advantageous solution in the server environment below the TOR.

# TECHNOLOGIES FOR IMPROVING PHOTONICALLY - ENABLED SOCKET-LEVEL I/O

Content provided by Zhihong Huang (HP Labs) and Clint Schow (UCSB)

There are a number of technologies in development that show promise in improving the efficiency and expanding the applicability of optical interconnects for datacenters and high-performance computing applications.

Currently, for short reach (< 50 m at 56 Gbps, 300 m at 10 Gbps), directly-modulated VCSELs with coarse wavelength division multiplexing (CWDM) are used in interconnects, but as energy demands tighten, dense wavelength division multiplexing (DWDM) may be a promising interconnect strategy. Implementing DWDM in integrated silicon photonics is an attractive solution, but there are a number of areas in the photonic link where silicon photonic technology must be improved to make it a viable option (see Figure 19). Here, we discuss modulation, detection, gain, and alternative paths to higher bandwidth.

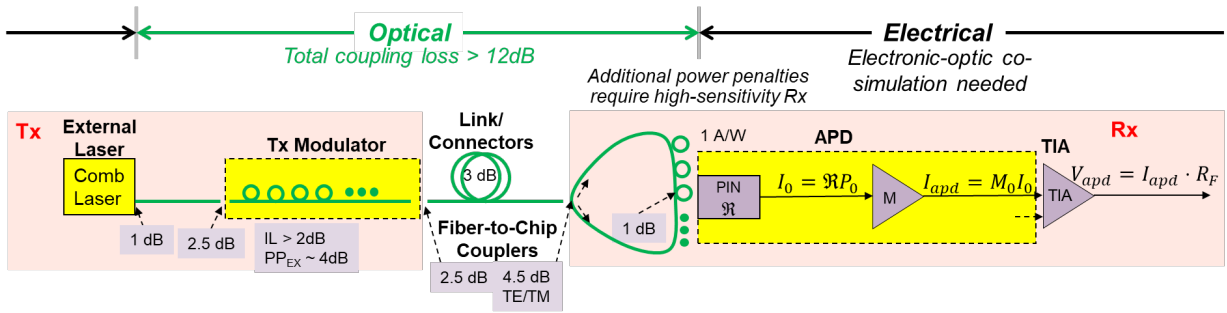

**Figure 19**: A notional optical link in silicon photonics.

On the transmitter side, micro-ring modulators are a good option for a DWDM link because they can perform both multiplexing and modulation functions in a single, compact device. However, further development is needed to achieve high speed and efficiency with low crosstalk between DWDM channels. They also require active tuning to ensure ring resonances are aligned with their assigned channels.

The detector on the receiver side needs both high speed and high responsivity with reduced bias voltage for compatibility with the computer power rail. Avalanche photodetectors (APDs) offer intrinsic gain to improve sensitivity, but have typically required high bias voltages that have made them inappropriate for interconnect applications. However, recent advances in SiGe APDs have demonstrated gain of 15, error-free operation at 25 Gb/s and a sensitivity of -16 dBm with a reverse bias of 10 V (54). Although the APD itself does not save a significant amount of power (receivers only consume a small portion of the power in an optical link), the improved sensitivity of the APD compared to a PIN receiver enables significant power savings in the lasers and transmitter (besides CMOS, most of the link power consumption is from lasers and modulators).

Gain within the link is closely related to the detector requirements. Specifically, adding a semiconductor amplifier (SOA) prior to the photodetector could significantly improve the sensitivity. Although amplified spontaneous emission (ASE) from SOAs will add noise to the optical link, it has already been shown that, at 2 km reach, an SOA results in a 7 dB sensitivity improvement compared to a 5 dB improvement from APDs (55). While this is promising, the Si platform does not have a built-in gain material, so a hybrid integration scheme needs to be developed. Recent examples use flip-chip bonding to integrate SOAs with Si photonics (56; 57), which have shown good performance, but are not particularly scalable.

Improvements in modulators and detectors are critical enabling technologies for increased interconnect bandwidth, but it is also important to consider the interconnect's data processing/encoding for improving bandwidth. Figure 20 shows the three degrees of freedom for increasing bandwidth: wavelengths/fibers (broadly described as channels), symbol rate, and bits/symbol. Moving along each of these axes has its own advantages and challenges.

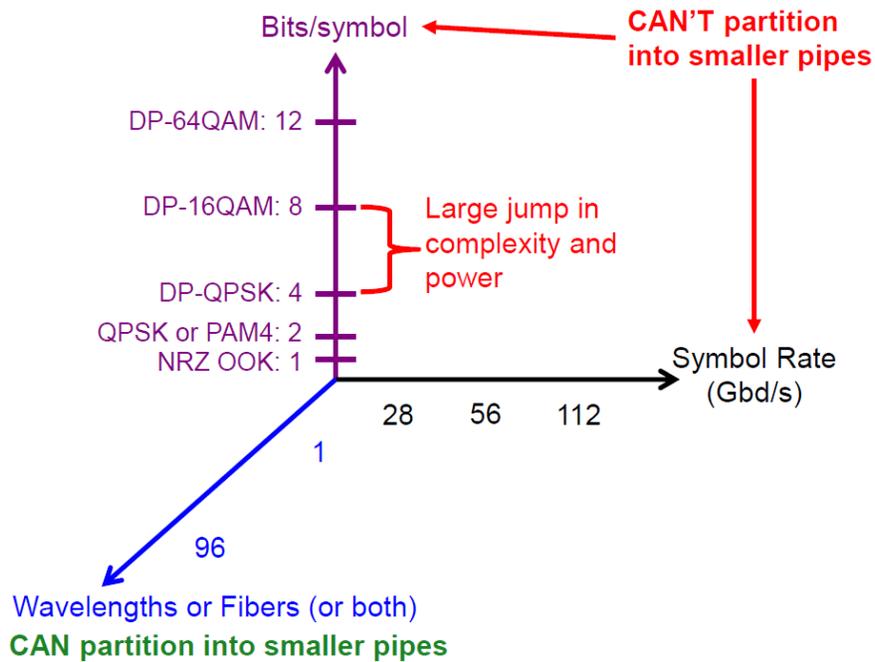

**Figure 20**: Pathways to higher optical link bandwidth. For reference, common modulation formats such as QAM (quadrature-amplitude modulation), QPSK (quadrature phase-shift keying), PAM (phase-amplitude modulation), and NRZ/OOK (non-return-to-zero / on-off keying) are depicted along the bits/symbol axis.

One new way to effectively increase the number of channels is spatial division multiplexing (with multicore fibers) and mode division multiplexing (with multimode fibers). Either type is projected to have a much higher spectral efficiency than single-mode fiber, especially over shorter reaches (58). On a Si photonic chip, modes can be multiplexed by careful design of the waveguides (59). Mapping the on-chip modes to those in the fiber requires specialized couplers. As this is a relatively new field of study, there is much room for performance improvement before this is an attractive interconnect method, but it could be an effective way to further increase the bandwidth of interconnects.

Advanced modulation formats while still employing direct detection are another path toward increasing interconnect bandwidth. These increase the data rate for a given baud rate, which is especially advantageous for bandwidth-limited channels. However, these modulation formats are more complicated, especially at the receiver, and they require forward error correction, adding latency (a particular problem for high-performance computing applications). Of the advanced modulation formats, pulse-amplitude modulation (PAM) is the least scalable. Although PAM-4 has half the baud rate of non-return-to-zero (NRZ) for the same data rate, it is problematic to move to more than four levels of pulse-amplitude modulation.

Thus, the next step for scaling with advanced modulation is coherent communication. In addition to higher spectral efficiency, coherent links also offer dramatically improved receiver sensitivity opening a path to larger optical budgets that can potentially enable other technologies such as all-photonic switching. Coherent transmitters are usually implemented via a Mach-Zehnder modulator (MZM). The receiver has to provide a reference signal to measure phase, and it has to handle polarization, dispersion, and frequency and phase offsets. In a conventional coherent receiver, this is accomplished with a free-running local oscillator (LO) and a polarization splitter, with digital signal processing (DSP) accounting for the other variations in the system. Alternatively, a phase-locked loop (PLL) can be employed. This makes the optics more complex, but no DSP, and potentially no error correction, is needed for quadrature phase-shift keying (QPSK), resulting in low latency. For higher-order advanced modulation such as QAM (quadrature-amplitude modulation), DSP is still required.

There are two platforms that are active areas of research for implementing coherent transmitter/receiver functionality in integrated photonics: InP and Si. InP has the advantage that, as an optically-active material, lasers for the source and LO can be integrated on-chip, and modulators are efficient. Low-loss polarization mux/demux functionality in InP does not yet exist. Si-based photonics, however, does include polarization mux/demux as well as passives and metal layers. Si is not an optically-active material, though, and thus laser sources are not directly integrated in this system. Optical isolators are a missing critical component in both platforms. Quantum dot lasers, which have substantially higher reflection tolerance compared to traditional quantum-well lasers may be promising for isolator-less systems.

When comparing the predicted energy efficiency of these two platforms, the dominant factor is the modulator efficiency. For a given bit error ratio (BER), InP modulators require a much lower drive voltage than typical Si-based modulators resulting in better energy-per-bit performance (see Figure 21). In designing a link, many additional power efficiency tradeoffs should be considered, including traveling wave versus segmented MZMs and modulator bias condition.

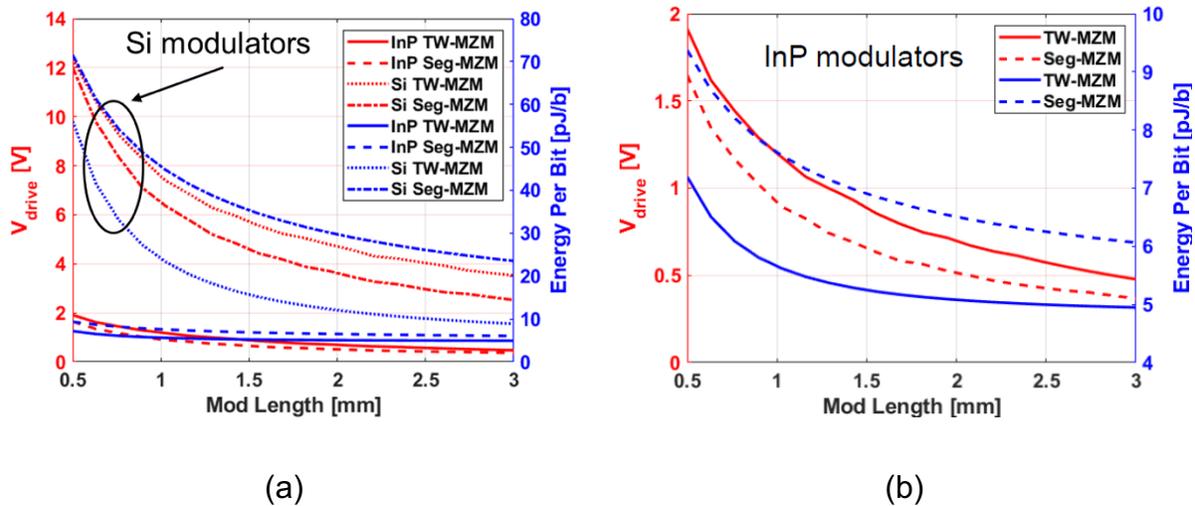

**Figure 21**: Simulated drive voltage and energy per bit with respect to modulator length for traveling-wave Mach-Zehnder modulators (TW-MZM) and segmented Mach-Zehnder modulators (Seg-MZM) at 50 Gbd and BER of $10^{-12}$. (a) Both InP and Si modulator performance. (b) Zoomed-in view of InP modulator performance.

Polarization multiplexing can be added to higher order modulation to effectively double the bandwidth per channel, and can be accomplished on-chip in a Si photonic platform (60). However, it significantly increases complexity. In some applications, it may be more appropriate to use DSP to implement standard multiple-input and multiple-output (MIMO) techniques for polarization demux.

In conclusion, for integrated photonics to meet the increasing bandwidth and efficiency needs of interconnects, many parts of the optical link must be improved. Link budgets are critical, but have often been overlooked, especially when considering the receiver technology. Coherent short-reach links show promise for low-power links; as a different optimization space compared to traditional telecom, there is a lot of room for exploration of new devices and architectures. As these technologies for improving bandwidth and efficiency are developed, close attention must be paid to their associated challenges and trade-offs.

# LIGHT GENERATION STRATEGIES

Content provided by Michal Lipson (Columbia University) and Di Liang (Hewlett Packard Labs)

Laser efficiency continues to be a major factor for scaling down the energy/bit of HPC optical interconnects. Gain material choices (GaAs/InP-based systems, QW/QD-based active regions) are generally driven by the need for direct-bandgap light emission. General requirements for low energy/bit lasers include low threshold current, high wall-plug efficiency, robust operation and reliability, and low cost. Additional requirements are driven by specific system architectures. Higher order modulation schemes (PAM4, QAM) may require narrow linewidth, low relative intensity noise (RIN), high power, and $\lambda$ tunability, while higher baud rates require low jitter, low chirp, and sometimes large direct modulation bandwidth (typically <40GHz). If parallel $\lambda$ channelization is desired, $\lambda$ control must be precise, and $\lambda$ spacing should be < 20 nm. Table 9 summarizes some tradeoffs for various light source technologies:

**Table 9:** (Direct) light source technology overview.

|  | Lasers on native substrates (InP, GaAs; QW, QD) | Heterogeneous integrated lasers (InP-, GaAs-on-Si; QW, QD) | Monolithically integrated lasers (InP-, GaAs, Ge-on-Si; QW, QD) |
|---|---|---|---|
| Pro | – Mature commercial technology<br>– Convenient integration in III-V PIC | – Maturing technology<br>– Convenient integration in Si PIC | – Potentially lowest cost bottom line<br>– Good thermal dissipation |
| Con | – Inevitable packaging difficulty if interacting with SiPh PIC<br>– High cost bottom line (InP-based) | – Proprietary volume fab technology<br>– Medium cost bottom line | – R&D stage<br>– Challenging to build a PIC |
| Product/demo | – VCSELs (0.85 –1 μm, 1.3-1.55 μm)<br>– DFB, DBR lasers (1.25 – 1.6 μm)<br>– Comb lasers (1.3, 1.55 μm)<br>– Others: FP, Ring, etc. | – DFB, DBR lasers (1.25 – 1.6 μm)<br>– Comb lasers (1.3, 1.55 μm)<br>– Others: FP, Ring, etc. | – DFB lasers (1.25 – 1.6 μm)<br>– Comb lasers (1.3 μm)<br>– Others: FP, Ring, etc. |

Coarse Wave Division Multiplexing (CWDM) and Dense Wave Division Multiplexing (DWDM) are distinctly different options for $\lambda$ channelization (wavelength-based parallelization) to increase aggregate I/O bandwidth. CWDM sources (4 to 8 single-$\lambda$ lasers, typically ~20 nm spacing) are often uncooled (saving energy), can be directly or externally modulated, and can be turned off individually when not being used (flexibility, energy efficient), however, they suffer fabrication and λ control issues as channel counts increase. DWDM sources (10s – 100s of λs from a single comb source, typically < 1 nm spacing) are appropriate for massively parallel wavelength architectures and advanced modulation formats. A comb laser can be a single element, or combs can be generated by optically pumping a specially designed optical cavity. Channel spacing is determined by device geometry, which can be controlled precisely in semiconductor fabrication, so individual wavelength control is not required to maintain constant channel spacing as long as temperature changes due to external environmental heating or self-heating are mitigated. For comb lasers yielding a flat output power spectrum across wavelengths, the 3-dB comb spectral width is typically 10 nm or less, so the total number of comb lines depends directly on channel spacing which can range from 10 GHz to 120 GHz. Close to 60 comb lines with 20 GHz spacing was demonstrated by UCSB in 2019 (61), and over 30 comb lines with 50 GHz spacing was reported by Fraunhofer HHI at OFC 2020 (62). Hewlett Packard (HP) Labs will soon publish a new result of over 200 comb lines with 35 GHz spacing from a single comb laser. Due to optical loss introduced by the saturable absorber in a comb laser cavity, 10% wall plug efficiency or above is the current state of the art. In contrast to single-λ lasers, for comb lasers there is always a tradeoff between the number of comb lines and individual comb line power. Presently it is challenging to achieve >3 mW power from each wavelength for more than ~24 comb lines/device, therefore a device outputting a uniform 32 comb lines with 3 mW/λ and 50-100 GHz spacing at 50°C represents a performance target for the near future, and would yield a very effective optical source for commercial use. Since comb lasers are always on regardless of wavelength utilization, they are only suitable for fully utilized DWDM applications from an energy efficiency perspective.

There are a number of notable demonstrations of directly modulated laser technologies in the research literature. Using CWDM VCSELs (λ ~ 1 µm, InGaAs/InP), HP Labs have achieved 112 Gb/s/λ using PAM-4 modulation and 4 λs to demonstrate a 400 Gb/s co-packaged solution for HPCs (63). Hitachi and Oclaro have demonstrated a directly modulated InGaAlAs/InP laser (λ = 1.3 µm) at 50 Gb/s on-off-keyed (OOK) (64). NTT Labs have demonstrated <1 pJ/bit 10 Gb/s directly modulated OOK lasers by monolithically integrating lateral InP p-i-n structures with photonic crystal waveguides (65). More recently in ECOC 2019 and OFC 2020, NTT Labs and II-VI/Finisar demonstrated specially designed directly-modulated lasers using photon-photon resonance to extend the 3-dB modulation bandwidth to over 100 GHz, enabling >300 Gb/s data transmission when combined with advanced modulation formats (66; 67; 68).

Quantum Dot (QD) materials present a "new" option for engineering of gain materials with high-temperature gain stability (i.e. temperature insensitive), wide optical gain bandwidth, low RIN, high tolerance to material defects, and low sensitivity to external optical feedback (particularly for single-λ lasers), all important factors for highly integrated applications. Commercial companies (QD Laser, Innolume) produce FP, DFB, and DBR lasers using a QD laser active region on GaAs that can operate from room temperature to 200 °C or higher (69; 70). Comb laser products generate wavelength combs with a few mW/λ output in the 1 – 1.3 μm range for external modulation applications.

Heterogeneous (or hybrid) integration has also proven successful with the advantage of combining disparate materials each with their own optimized properties, with the disadvantage of fabrication complexity and potential reliability challenges. Integration is achieved through bonding traditional optical gain materials (example InGaAs/InP) with large-scale silicon-on-insulator (SOI) wafers. Intel's commercial 100G PSM4 and CWDM4 QSFP transceivers employ Si photonics hybrid integrated with InP-based lasers at λ ~ 1.3 μm, and have a proven track record of performance and reliability (71). UCSB and Caltech have demonstrated single-λ lasers at 1.55 μm using similarly hybrid integration schemes with low-loss Si/SiO$_2$ micro-ring resonators, with UCSB achieving laser linewidths as low as 220 Hz (72; 73; 74). HP Labs have achieved directly modulated hybrid-integrated QD lasers (InAs/GaAs QD on Si micro-rings) with <2 pJ/bit at 15 Gb/s (75), 25 Gb/s using optical injection locking techniques, and 15-λ 100 GHz spaced comb lasers with RIN performance rivaling external cavity lasers (76). They have also demonstrated extremely energy efficient and high-speed electrical functionalities by sandwiching a dielectric layer (native or deposited oxide) at the bonding layer to create low-leakage MOS capacitors (77). This structure induces plasma dispersion effects for low-energy high-speed filter tuning and frequency modulation (1.6 nm/pW) in micro-ring structures.

Where possible, monolithic integration is most desirable for fabrication simplicity, cost, and reliability. Several leading research groups (UCSB, University College London, University of Tokyo) have monolithically grown III-V QD structures on Si, demonstrating reduced impacts of dislocations (relative to QW structures) resulting from significant lattice mismatch (78; 79). Monolithic QD structures on Si have been successfully used for DFB arrays (80), microring and mode-locked lasers (61), with reliability (Time to Failure) stead-ily improving over time. Combining the best of both worlds (monolithic and heterogeneous integration), researchers at NTT and HP Labs have demonstrated high-quality lasers fabricated by combining InP-based epitaxy with wafer bonding III-V material to SOI (81; 82). This process will allow III-V epitaxy on much larger-scale silicon substrates of 200 or 300 mm to reduce epitaxy cost. A 10 Gb/s directly-modulated laser by NTT on this platform demonstrated an extremely low threshold of 24 uA and record high energy efficiency of 7.3 fJ/bit recently (83). In addition, researchers are pursuing similar techniques to achieve monolithic integration (on a single chip) of critical optical functions necessary in a transceiver (laser, optical modulator, optical preamplifier, photodetector).

Table 10 summarizes design trade-offs for on-chip laser sources:

**Table 10**: Summary of design trade-offs for on-chip laser sources **(84; 85; 78)**.

| On-chip integration | Hybrid Integration or heterogeneous growth of the gain medium on chip |
|---|---|
| **Gain Media** | III-V SOAs, III-V Quantum dot / Quantum Well structures, Ge-Si, Erbium doped waveguide structures |
| **QD/QW sources** | Pro: High optical powers, O-band operation<br><br>Con: Laser ridge WG (1-2 x 4-10 um) -> mode mismatch with planar WG structures for hybrid / heterogeneous integration |
| **III-V SOA** | Pro: Mode matching w/ planar waveguides minimizes coupling loss for C/L band operation<br><br>Con: Lower optical power than QD/QW |
| **Ge / Er+ based sources** | Con: Lower optical powers inadequate for coherent telecom or datacom<br><br>Con: Material Challenges need to be addressed for on-chip lasers |

Review:
Liu et al., Photon. Res. 3, B1-B9 (2015)
Zhou et al., Light: Science & Applications (2015)
Komljenovic et al., J. Light. Technol (2016)

In addition to the limitations imposed by these tradeoffs, laser efficiency typically increases with laser power, driven in part by the need (and energy/bit cost) for temperature stabilization (i.e. TE coolers and power supplies).  For example, electrical-to-optical energy conversion efficiencies as high as 30% can be achieved for diode laser output powers in excess of 10 mW, whereas efficiency drops sharply as output power approaches 1 mW.  This has motivated significant work in developing high-power external wavelength-comb generating lasers.

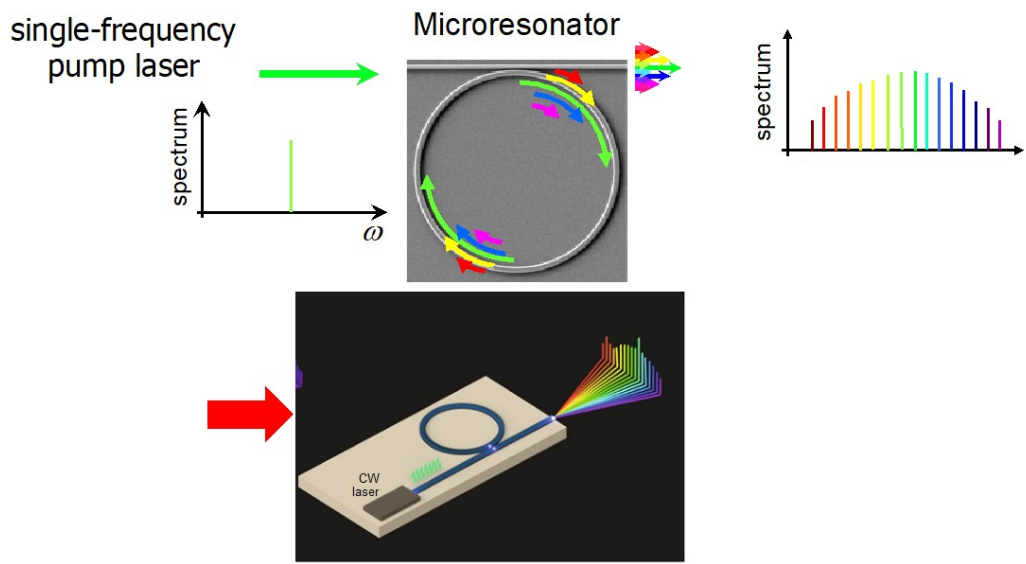

**Figure 22**:  Frequency comb generation in a micro-ring.

Many approaches involve driving an optical resonator (micro-ring) into a nonlinear regime using a high-power single-frequency pump laser, exploiting soliton behavior to generate a frequency (wavelength) comb.  By carefully balancing linear (resonant) and non-linear effects, researchers have generated frequency combs with 100s of wavelengths.  Adding optical amplification (often integrated SOAs) increases the power/frequency to useful levels for external modulation and data transmission.

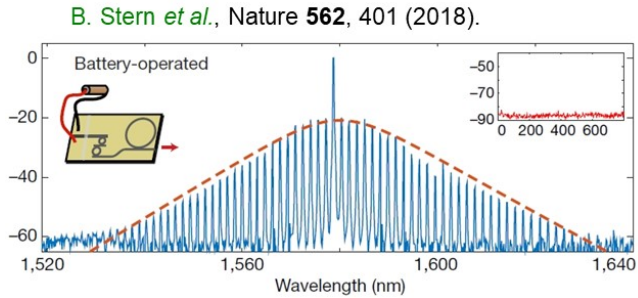
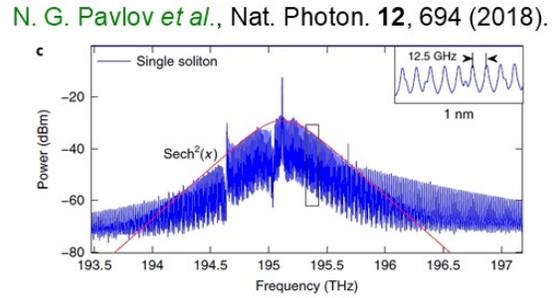

**Figure 23**: Experimental results for frequency comb generators **(86; 87)**.

Unfortunately, the power output of each comb line in the configurations of Figure 22 and Figure 23 varies as much as 40 dB across the wavelength band (decreasing to either side of center), meaning efficiency is strongly wavelength dependent, and overall efficiency (energy/bit) is adversely impacted due to the requirement to meet system margin requirements at all wavelengths, not just the most energetic ones. Combining two ring resonators into coupled configurations enables some flattening of the resulting output frequency comb spectrum and power. Pump-to-comb (optical-to-optical) conversion efficiencies as high as 40% have been demonstrated in this matter.

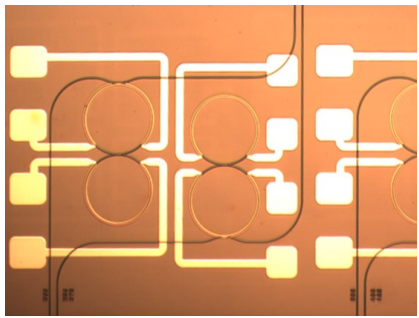
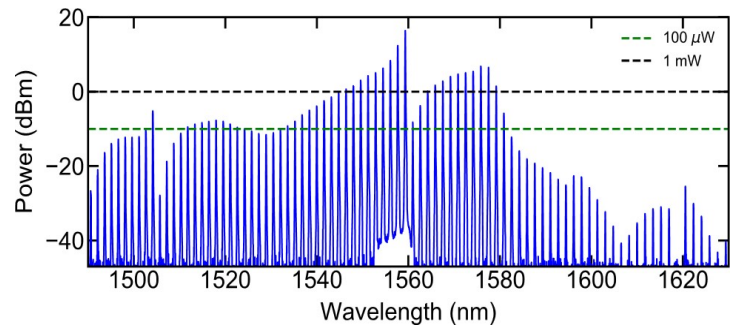

**Figure 24**: Frequency comb generation with coupled micro-ring resonators (left) for flattened output wavelength spectrum (right).

In summary, there is a diverse set of technologies available for light generation to meet the demanding requirements of optical interconnects in HPC systems. For low-wavelength counts where direct modulation is desired, CWDM VCSELs can provide high modulation speeds (baud rates) for OOK and PAM4 modulation formats. For larger wavelength counts, external comb lasers provide potentially 100s of wavelengths with narrow linewidths for higher-order external modulation schemes like QAM. QD technologies show great promise for both laser and optical amplification applications, primarily due to their temperature insensitivity, broad optical bandwidth, and reduced sensitivity to dislocation defects. Monolithic integration is always most preferable, but often not possible due to the disparity of materials required for an end-to-end optical link, therefore, advances in heterogeneous/hybrid integration will likely continue as key drivers in next-generation technologies. No single technology is likely to dominate the application space in the near future. Rather, specific system-level requirements for HPC optical interconnects will drive selection of the appropriate technologies and tradeoffs for light generation.

# PHOTONIC INTEGRATED CIRCUIT FOUNDRY OVERVIEW

Content provided by Edward Preisler (Tower Semiconductor)

One of the biggest reasons for the low cost of electronics is the extensive foundry infrastructure for CMOS fabrication, which makes it simple to make standard CMOS circuits. Photonic integrated circuits (PICs), as a less mature technology, do not benefit from the same scale of foundry infrastructure, and this is a contributing factor to higher costs. We discuss the characteristics of PICs that differentiate them from CMOS for a foundry-type process, the current status of PIC foundries, and the biggest hurdles for the establishment of a more universal PIC foundry infrastructure.

Commercial silicon foundries span a wide range of customization, trading off between design flexibility and economy of scale advantages. The least flexible foundries are for digital CMOS, which do not allow any customization, and customer intellectual property (IP) is contained in the way the foundry's standard blocks are connected. In contrast, completely customized foundries, such as those for micro-electro-mechanical systems (MEMS), will even adjust the process itself on a customer-by-customer basis. Currently, PIC fabrication would fall most appropriately in the middle: the fabrication process and some modules are standardized, but the customer directly defines the layout of all layers.

Using commercial CMOS foundries can be beneficial to PIC customers in a few ways. The existence of process design kits (PDKs) can dramatically reduce design effort and time to market. In addition, the statistics built up by the foundry can give a better understanding of the origin of yield issues. Finally, there is an abundance of capacity; most large CMOS fabs could handle the world's entire current silicon photonic demand without significantly perturbing their existing business. At the same time, there are challenges that come with using a commercial foundry. Silicon photonic tape-ins are notoriously difficult for verification software, and customers and the fab must compromise on what rule violations are acceptable to waive. Traditional in-line process controls used for CMOS processes are suitable for controlling SiPh processes but traditional electrical wafer acceptance test methodologies used for CMOS processes are not sufficient for SiPh processes. Wafer-scale optical testing of needs to be developed to close this gap.

Given the challenges associated with making PICs in traditional large-scale commercial CMOS foundry processes, compromises need to be made by the foundry in order to be able to manufacture PICs while still leveraging the advantages of large-scale manufacturing. It is straightforward to develop a PDK of passive devices and photodiodes that satisfies the needs of all customers. However, there is currently too wide a range of customer needs to provide broadly applicable modulators, fiber-to-chip couplers, and 3D wafer manipulations.

Table 11: Existing PIC foundries (semiconductor foundries in orange).

| Foundry | Capacity (wafers/month) | Platform | Business Model |
|---|---|---|---|
| AIM - US | <100 | 300 mm - thin SOI | Gov't / Industry consortium, multi-project wafers (MPWs) only |
| AMF - Singapore | 100s | 200 mm - thin SOI | Pure-play (PP) foundry, prototyping and low-volume production |
| GF - Singapore / US | 1000s | 200/300 mm – thin SOI | PP foundry but partnership-only model for silicon photonics? |
| IHP - Germany | 100s | 200 mm - thin SOI | Gov't / Industry consortium, low-volume production |
| IMEC - Belgium | <100 | 200/300 mm – thin SOI | Industry consortium, MPWs only |
| LETI - France | <100 | 200 mm – thin SOI | Gov't / Industry consortium, MPWs only |
| Sandia - US | <100 | 150 mm – thin SOI | Gov't lab, MPWs only |
| Smart Photonics – Netherlands | 100s | 100 mm – InP | PP InP PIC process foundry |
| Skywater - US | 100s | 200 mm – thin SOI | PP foundry, medium volume? |
| Tower Semiconductor - US | 1000s | 200 mm – thin SOI | PP foundry, high volume |
| TSMC - Taiwan | 1000s | 300 mm – thin SOI | PP foundry but partnership-only model for silicon photonics? |
| VTT - Finland | <100 | 150 mm – thick SOI | Gov't / Industry consortium, low-volume prototyping |

Nonetheless, the ecosystem of silicon photonic foundries has been growing. Table 11 is a list of existing players. Typical foundries offer PDKs with various waveguide shapes, splitters, couplers, transitions, photodiodes, and phase shifters. Figure 25 shows the reported/advertised performance specifications of devices offered by a number of different foundries. In general, they all have comparable performance levels. In addition to these typical PDK items, there is some movement toward integration services. CMOS integration, which is already offered by a couple of foundries, may result in some performance gains, but there are issues with yield, cost, and cycle time. III-V materials are attractive for active devices, but integration methods are still under development. Laser die attachment seems to be a decent "brute force" solution, but other heterogeneous III-V integration paths are being pursued. It is also possible to implement the entire PIC process in InP in a foundry process, but waveguides are much larger and lossier than silicon.

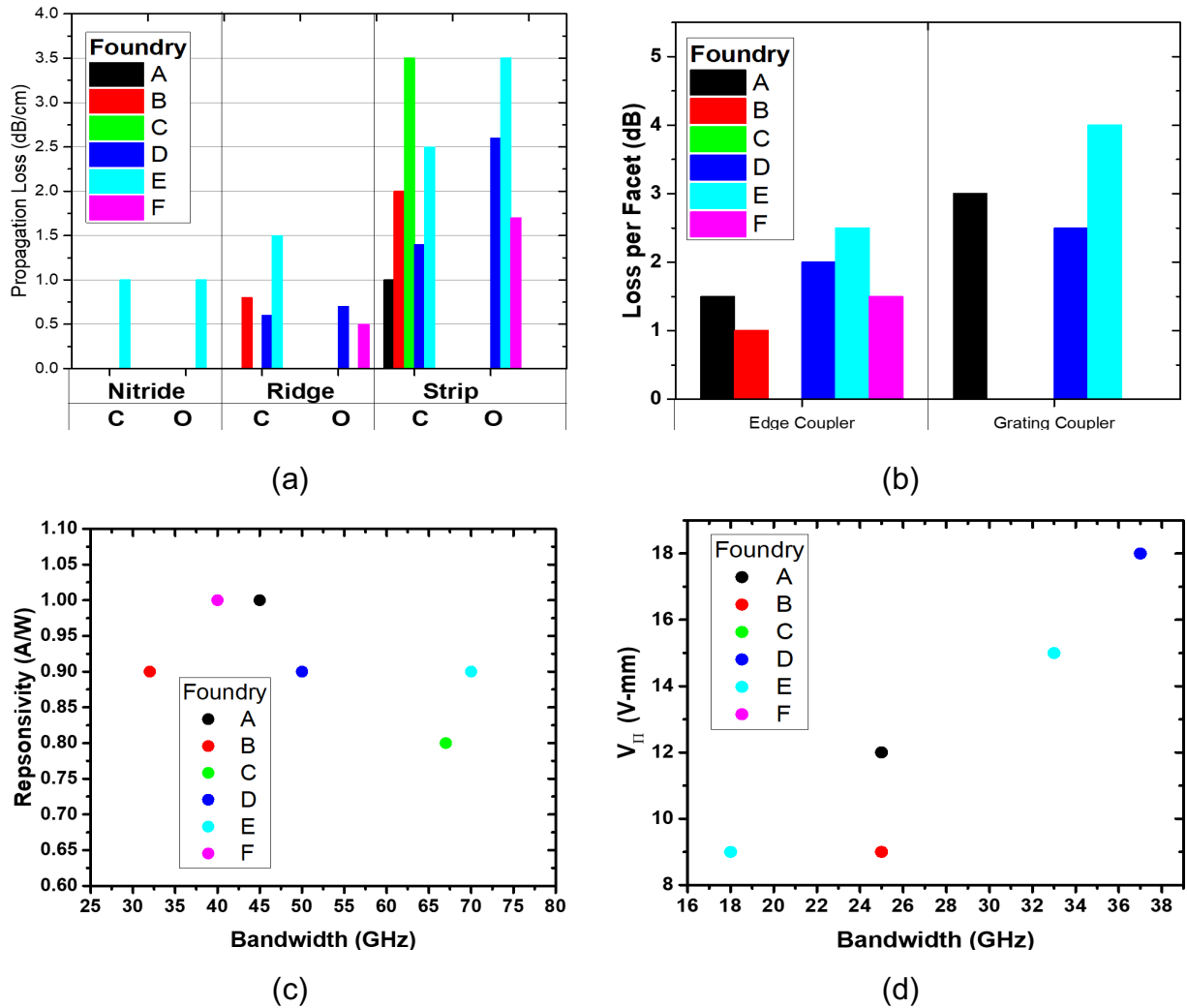

**Figure 25**: Some baseline foundry silicon photonics offering specifications. (a) Waveguide propagation loss in both the C and O bands. (b) Fiber-to-chip coupling loss. (c) Germanium photodiode performance. (d) Mach-Zehnder modulator performance.

Unlike with CMOS, packaging is an aspect of PICs that must be considered as a part of the foundry process. The PIC layout and processing must anticipate the singulated die packaging scheme. This would include V-grooves and edge couplers, laser trenches and pedestals, and isolated through-silicon-vias. The wide range of packaging techniques is yet another source of customer-specific needs that has to date made it difficult for a PIC foundry to have a standardized process.

As a whole, the number of PIC foundry services is growing, but there are many aspects that need further development to achieve similar broad applicability, reliability, and low cost as CMOS foundries. Areas ripe for technological advancement include yield analysis, in-line process monitoring techniques, and standardization of fiber-to-chip coupling and interfacing with packaging.

# AIM PHOTONICS: TEST, ASSEMBLY, AND PACKAGING FACILITY
### Based on materials and presentation provided by Ed White (AIM Photonics)

Packaging and testing are critical parts of implementing integrated photonic links in real-world applications, and standard processes addressing this are under development. One facility that is working on developing a suite of capabilities for integrated photonic packaging is the American Institute for Manufacturing (AIM) Integrated Photonics Test, Assembly, and Packaging (TAP) facility. The challenges they are addressing give a window into the general state of the field of photonic packaging.

The AIM TAP facility is focused on both wafer-scale and chip-scale capabilities with both optical and electrical interfaces. At the wafer scale, services include optical, radio frequency (RF), and DC testing, PIC and interposer metallization, die bonding of lasers and PICs, and singulation of dies. Chip-scale capabilities also include fiber attachment and surface mount assembly.

In order to provide these services, the AIM TAP facility has developed new processes and tools specific to photonics packaging. As there does not yet exist an industry-standard suite of PIC packaging and testing tools, AIM TAP often repurposes existing electron-ics processing tools. For example, a flip-chip tool is used for bonding PICs with interposers, and electrical testing is accomplished using traditional wafer-scale testers. At the same time, some tasks do not have an analog in the electronics industry, so more customized solutions are necessary. Examples include fiber attachment and RF and photonic testing tools.

Because PIC packaging is a newer field, there are a number of areas with significant room for improvement. One of the biggest challenges AIM TAP has seen in the packaging process is optical coupling. Achieving better than 1 dB loss at a single facet quickly and consistently is very difficult, requiring active alignment. For more scalable, low-cost devices, fiber arrays (not just single fibers) will need to be attached with passive alignment – a capability that has not yet been demonstrated. Higher density electrical I/O is also needed; consistent 75 µm bump pitch is currently possible, but future applications will require much smaller pitch size.

Overall, PIC packaging and testing is a complex process involving electrical, RF, and photonic interfaces. In order to be successful, the packaging process must be low-cost and reliable. More work is needed to develop and standardize techniques and tools that satisfy these needs.

# ULTRA-ENERGY EFFICIENT MULTI-TERABIT PHOTONIC INTERCONNECT PHOENIXSIM ENVIRONMENT

Content provided by Keren Bergman (Columbia University)

A robust development and supply chain for integrated photonics supporting high-speed socket-level I/O interconnects in High Performance Computing (HPC) systems will require sophisticated simulation tools that can accurately predict device and system level performance. In particular, device-level simulations for a diverse set of external lasers (DFB, comb lasers, VCSELs), waveguides/couplers, micro-ring modulators, photodetectors, and all-optical switches must produce accurate performance metrics (laser linewidth and RIN, waveguide and ring insertion loss, modulator and photodetector speed, optical switch and inter-modulation crosstalk) that can feed directly into system-level simulations and metrics (optical power budget, energy/bit, I/O bandwidth). In addition, these simulation tools must support link optimization where design parameters like wavelength channel spacing, modulation scheme (OOK, QAM, PAM-4), and filter array geometries (micro-rings, arrayed waveguide gratings, interleavers) can be varied to achieve the lowest energy/bit (typically <1pJ/bit) and maximum data throughput possible.

PhoenixSim is a state-of-the-art modular simulation environment from Columbia University's Lightwave Research Laboratory that couples Electronic Design Automation (EDA) tools like Semiconductor Device Modeling, Electrical Modeling, and Electronic Circuit Design/Layout with Photonic Design Automation (PDA) tools like Photonic Component Modeling, Photonic Link Design, Photonics System/Networks, and Layout Implementation. A good example demonstrating PhoenixSim capability is a Si photonic DWDM link that employs micro-rings for both wavelength-selective OOK modulation (transmitter) and pre-photodetection wavelength selection (receiver).

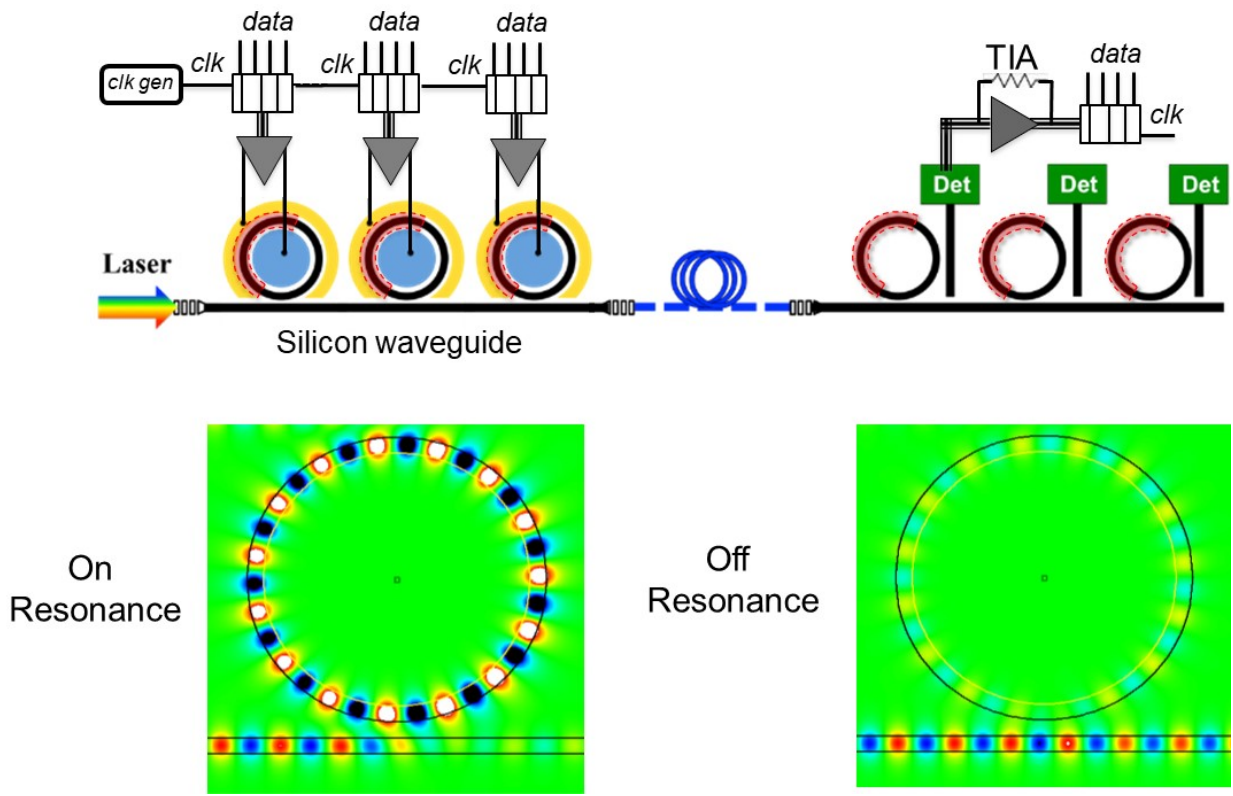

**Figure 26**:  Basic Si photonic link for PhoenixSim simulation.

In this example, for OOK modulation PhoenixSim accurately predicts key performance parameters for the left-hand transmitter in Figure 26 (optical eye-diagram extinction ratio, insertion loss, rise/fall times, channel crosstalk) by combining analytical models for the electrical drive signal at data rate, micro-ring electrical frequency response, micro-ring resonance thermal tuning response, and laser source comb frequencies/wavelengths (number of channels and spacing) with degradations due to free carrier effects (absorption, dispersion), and physical layer structure (ring radius, input/output gaps) to optimize the design parameters for minimum link power penalty (dB).

PhoenixSim directly supplements established electronic design tools like SYNOPSIS, integrating predictions of electro-optic performance with more traditional tools for designing transmitter CMOS circuitry, including clock dividers/multipliers, electrical drivers, and serializers/de-serializers. On the receiver (right hand side of Figure 26), PhoenixSim quantifies adjacent-wavelength crosstalk for each micro-ring de-multiplexer, predicting the power penalty (dB) one can expect on a Bit-Error-Rate (BER) curve versus received power for a variety of wavelength/channel spacings. The ultimate outcome of these simulations is full-link end-to-end design tools, where design parameters can be varied to optimize for top-level system performance metrics like lowest energy/bit and highest total I/O throughput. Taking this a step further, higher-level architectures can be evaluated for overall performance, an example being performance comparisons of DWDM point-to-point links using single-channel micro-ring modulators and wavelength demultiplexer/drop ports versus parallel architectures enabled by odd/even wavelength interleavers/de-interleavers. Finally, PhoenixSim integration with foundry PDKs and SYNOPSIS enables direct generation of GDS mask layouts once design optimization has been achieved.

In an important example, Columbia's Lightwave Research Laboratory has used PhoenixSim to explore the tradeoffs between use of various interleaver schemes and per-wavelength/channel bit rates to achieve 800Gb/s DWDM links. Specifically, comparing OOK 25Gb/s*32-Channels (8$\lambda$s interleaved by 4) with 10Gb/s*80-Channels (10$\lambda$s interleaved by 8), and considering both simple Mach-Zehnder (MZ) interleavers as well as more advanced resonator-assisted MZ and multi-stage interleaver designs, they find that the 10Gb/s*80-Channel design can reduce energy/bit by ~1/3 from 3.2pJ/bit (for the 25Gb/s*32-Channel design) to 2.2pJ/bit.

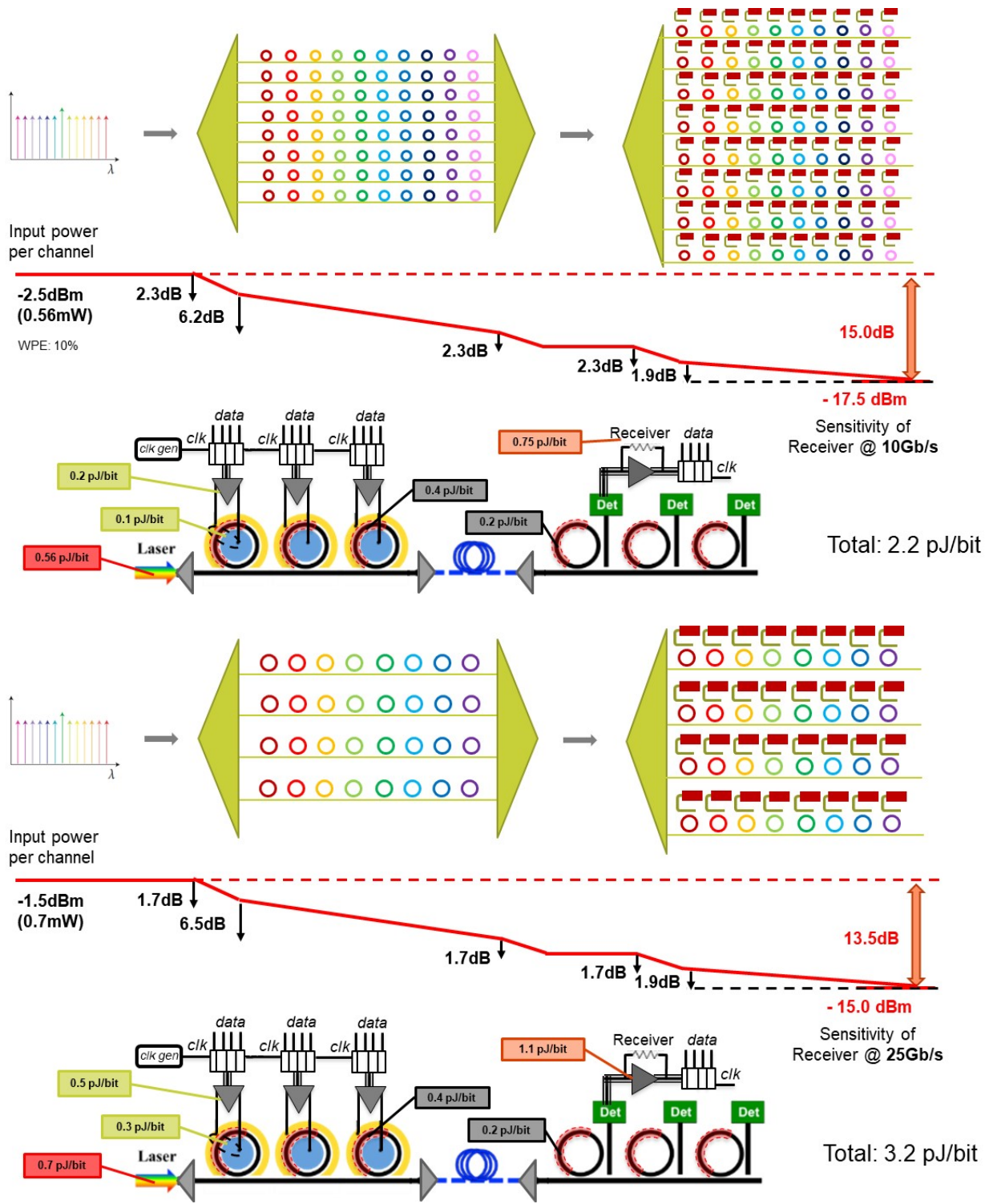

**Figure 27**: PhoenixSim power consumption analysis results comparing 10Gb/s*80-channel design (upper diagram) to 25Gb/s*32-channel design (lower diagram).

Detailed parameters used in the simulation (demux wavelength response, photodiode responsivity, modulator thermally-tuned resonance response, modulator bandwidth and equivalent circuit model, and component insertion loss) were independently confirmed experimentally for accuracy of results. Figure 28 summarizes the detailed breakdown of energy/bit penalty for each part of the link, demonstrating that primary energy savings (on a percentage basis) at 10Gb/s/channel came due to the significantly decreased pJ/bit cost of driver electronics.

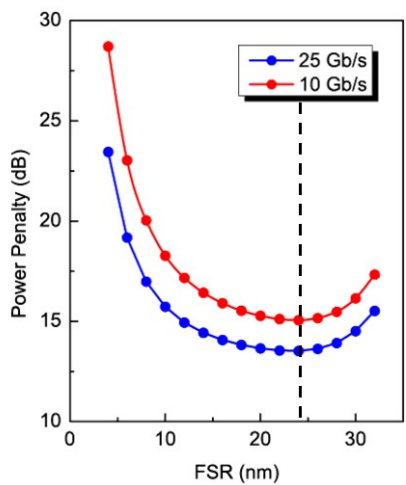
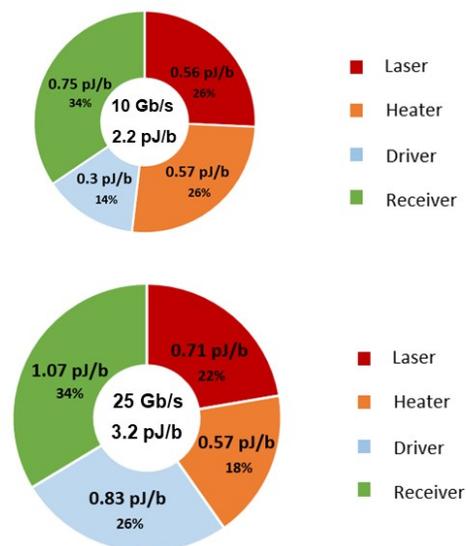

**Figure 28**: Results of PhoenixSim power consumption analysis broken down by major subsystems of the link.

In summary, PhoenixSim clearly demonstrates that detailed device performance parameters for a diverse set of photonic link building blocks (lasers, modulators, multiplexers/demultiplexers, photodetectors, drive/amplification electronics) enable simulated link optimization where design parameters like device dimensions, wavelength channel spacing, modulation data rate, and filter array geometries (micro-rings, interleavers) can be systematically varied to achieve the lowest energy/bit and maximum data throughput possible before chips are fabricated. Once optimization is complete, PhoenixSim (properly integrated with other design tools and specific PDKs) directly generates mask layouts for foundry fabrication of components and full up links.